\documentclass[11pt,a4paper]{article}
\pdfoutput=1
\usepackage{jcappub}
\usepackage{graphicx}% Include figure files
\usepackage{dcolumn}% Align table columns on decimal point
\usepackage{bm}% bold math
\usepackage{amsmath}
\usepackage{epstopdf}
\begin{document}
\title{Towards accurate cosmological predictions for rapidly oscillating scalar fields as dark matter}
\author[a]{L. Arturo Ure\~na-L\'opez}
\author[a,b]{and Alma X. Gonzalez-Morales}%
 
\affiliation[a]{%
Departamento de F\'isica, DCI, Campus Le\'on, Universidad de
Guanajuato, 37150, Le\'on, Guanajuato, M\'exico.}
\affiliation[b]{Consejo Nacional de Ciencia y Tecnolog\'ia,
Av. Insurgentes Sur 1582. Colonia Cr\'edito Constructor, Del. Benito   Juárez C.P. 03940, M\'exico D.F. M\'exico}
\emailAdd{lurena@ugto.mx}
\emailAdd{alma.gonzalez@fisica.ugto.mx}

\date{\today}

\abstract{
As we are entering the era of precision cosmology, it is necessary to
count on accurate cosmological predictions from any proposed model of
dark matter. In this paper we present a novel approach to the
cosmological evolution of scalar fields that eases their analytic and
numerical analysis at the background and at the linear order of
perturbations.  The new method makes use of appropriate angular
  variables that simplify the writing of the equations of motion, and
  which also show that the usual field variables play a secondary role
in the cosmological dynamics. We apply the method to a scalar field
endowed with a quadratic potential and revisit its properties as dark
matter. Some of the results known in the literature are recovered, and
a better understanding of the physical properties of the model is provided. It
is confirmed that there exists a Jeans wavenumber $k_J$, directly
related to the suppression of linear perturbations at wavenumbers
$k>k_J$, and which is verified to be $k_J = a \sqrt{mH}$. We also
discuss some semi-analytical results that are well satisfied by the
full numerical solutions obtained from an amended version of the CMB code
CLASS. Finally we draw some of the implications that this new
treatment of the equations of motion may have in the prediction of
cosmological observables from scalar field dark matter models.}
%\pacs{98.80.-k,95.35.+d,98.80.Jk}

\maketitle

\section{Introduction \label{sec:introduction-}}
The most recent observational results from the Cosmic Microwave Background
(CMB) give strong support to the idea of a benchmark model
of the Universe that is consistent with a variety of observational
data\cite{Ade:2015lrj}. It is remarkable that this standard model,
well known as the $\Lambda$CDM model, is at its core just described by
six numbers with a true cosmological context; the rest of the
physical quantities and constants, which are important for a complete
understanding of the high and low-energy processes in the evolution of
the Universe, can be fixed and measured by other methods at
terrestrial laboratories.

One of such cosmological numbers is particularly important to
understand the formation of cosmological structure: the
actual amount of cold dark matter (CDM), $\Omega_c h^2 \simeq 0.119$.
Dark matter represents about $27\%$ of the total matter budget. The base
model for CDM is as simple as it gets: a  pressureless fluid of 
particles that have no direct, weak at most, interaction with other matter fluids apart from the gravitational one. These two basic properties are
enough to provide a complex picture of structure formation through
hierarchical assembling that fits data at different
scales\cite{Vogelsberger:2014dza,Genel:2014lma} (see
also\cite{Bertone:2016nfn} for a historical account of dark matter). A
detailed model of CDM is necessary, but so far observations seem not to be very
demanding in that regard\cite{Yang:2015cva,Angrick:2015gta,Barranco:2013wy,Cuoco:2015rfa},
and a direct detection of the dark matter particle in terrestrial
laboratories has so far produced only null results, though significant
progress is being
made \cite{Amole:2015pla,Gluscevic:2015sqa,Ringwald:2015lqa}.

A satisfactory answer to the CDM riddle have to be complemented with a convincing theoretical model that give us a comprehensive understanding of its properties. Although there is quite a large number of proposals, here we will focus in a type of model that has been studied for more than two decades: scalar field dark matter (SFDM)\cite{Sin:1992bg,Ji:1994xh,Matos:2000ng,Matos:2000ss,Matos:2008ag,Sahni:1999qe,Marsh:2010wq,Marsh:2013ywa,Woo:2008nn,Hwang:2009js,Park:2012ru,Beyer:2014uja,Li:2013nal,Urena-Lopez:2013naa,UrenaLopez:2008zh,Lundgren:2010sp,Aguirre:2015mva,Amendola:2005ad}. The
key concept in this model is that the dark matter properties can be
described by the dynamics of a scalar field with a probably
fundamental origin, and a very tiny mass of around $10^{-22} \, {\rm
  eV}$,  as suggested by different cosmological and astrophysical
  constraints. Cosmological scalar fields are better known in the
context of dark energy models because many proposals are able to
provide an accelerating expansion of the Universe at late
times\cite{Copeland:2006wr}; however,  for dark matter models it has
been widely shown that any successful model must be at the end
related, somehow, to the simplest possible case: a scalar field
potential with a parabolic shape around a minimum (see
\cite{Marsh:2015xka,Magana:2012xe} for comprehensive reviews).

The typical form of the scalar field potential in SFDM models is:
\begin{equation}
  V(\phi) = (1/2) m^2 \phi^2 \, , \label{eq:0}
\end{equation}
where $\phi$ is a real scalar field, and $m$ is its mass
scale. As we said before, other potentials exist for SFDM models, but
their common feature is that they all can be approximated by the
functional form~\eqref{eq:0} once they enter the dark matter
phase, see for
instance\cite{Matos:2000ss,Matos:2009hf,Sahni:1999qe,Beyer:2014uja,Amendola:2005ad,Marsh:2013ywa}
 (although higher order corrections to the potential can play an
  important role in other high and low-energy
  processes\cite{Marsh:2015xka,Li:2013nal,Barranco:2011wq,Chavanis:2016dab}). The
resultant behavior of scalar fields as a pressureless fluid
appears from their rapid oscillations around the minimum of the
potential~\eqref{eq:0}, which is a well-known
fact\cite{Turner:1983he,Ratra:1990me,delaMacorra:1999ff}. For rapid
oscillations we mean that the period of the field oscillations,
represented by the time scale $m^{-1}$, is
much shorter than the typical Hubble time $H^{-1}$ for the expansion
of the Universe, where $H$ is the Hubble parameter. Because the Hubble
parameter is a decreasing function of time, the ratio $m^{-1}/H^{-1}$
quickly becomes very small for realistic models, and then accurate
numerical methods are required for a precise evolution of the
equations of motion\cite{Magana:2012xe}. This difficulty has been
overcome somehow by the use of approximate formulas, see for
instance\cite{Amendola:2005ad,Hlozek:2014lca,Hwang:2009js,Beyer:2014uja,Ratra:1990me,Cembranos:2015oya},
that are intended to match as smoothly as possible the transition
towards the pressureless phase of the scalar field dynamics (see also\cite{Alcubierre:2015ipa}).

As for any dark matter model, the properties of SFDM models must also
be connected to the formation of structure in the Universe, and then
the same physical properties that allow a consistent evolution at
cosmological scales must also be appropriate for the development of
realistic galaxy halos. The latter have been also extensively studied
in the literature for SFDM
models, see for
instance\cite{Martinez-Medina:2015jra,Martinez-Medina:2014hka,Robles:2015gwa,Robles:2014ysa,Robles:2013dfa,GonzalezMorales:2012uw,Diez-Tejedor:2014naa,Robles:2013dfa,RindlerDaller:2011kx,Rindler-Daller:2013zxa}
and references therein.
More recently, complete  wave-like simulations have shown the similarities
of SFDM with CDM\cite{Schive:2014dra,Woo:2008nn} at large scales, although most of these studies have been done under the non-relativistic
approximation. One another quantity that plays an important role here
is the mass Power Spectrum (PS), which represents the aggregation properties
of matter at different scales, whose measurement are for now in good
agreement with the theoretical expectations of the $\Lambda$CDM
model if the scalar field mass is large enough. Nonetheless, the
mass PS is sensitive to the physical properties of the
dark matter particles, and there may be noticeable consequences in the
history of galaxy formation if the dark matter component is not
strictly cold, as has been shown in\cite{Enqvist:2015ara,Boehm:2001hm,Marsh:2013ywa,Mead:2013jta,Mead:2014gia,Lombriser:2015axa}.

Given the interest in SFDM models in the recent literature, it is
desirable to have an improved handling of their cosmological dynamics,
for variables in the background and in linear perturbations,
even if eventually one has to use numerical methods to resolve them
completely. Along these lines, the aim in this paper is two
fold. First, we would like to present a new approach for the solution
of the cosmological equations of SFDM, so that we can have a proper
understanding of the subtleties in their evolution. This will be
achieved through the use of the tools of dynamical systems (for some
pedagogical presentations see\cite{UrenaLopez:2006ay,Faraoni:2012bf,Garcia-Salcedo:2015ora}),
together with appropriate changes of variables that are more than
suitable to exploit the intrinsic symmetries of the equations of
motion. The techniques developed here can be easily extended to
different cosmological settings in which scalar fields may be
involved. Second, we will analyze the mass PS of SFDM models for different masses of the SFDM particle draw some implications on other observables such as the mass function, in a approximate way since the scales of interest are  beyond the realm of linear perturbations.

A brief summary of the paper is as follows. In Sec.~\ref{sec:scal-field-dynam} we explain the conversion of the equations of motion for a scalar field dominated Universe into an
autonomous dynamical system, where the latter is written in terms of a
kinetic variable and two potential variables. Taking into account the
symmetries of the system of equations, we propose a change to polar
coordinates that helps to convert them into a more manageable
2-dimensional dynamical system that can be solved numerically without
any by-hand manipulations. As we shall show, the new method is able to
follow the evolution of the system from the early slow-roll regime to
that of rapid oscillations with high precision and minimal numerical
effort.

In Sec.~\ref{sec:line-axion-pert}, we apply similar mathematical
manipulations to the equations of motion of linear perturbations, so
that their evolution is also given in terms of a 2-dimensional
dynamical system. All the new equations, for the background and for
linear perturbations, will be put into an amended version of
CLASS\cite{Lesgourgues:2011re}, which will be used to calculate all
relevant cosmological quantities. We will clearly identify, within the
new system of equations, the so-called Jeans scale of SFDM. As a side
result, the velocity gradient is also calculated to verify the
appropriateness of the synchronous gauge for the calculation of linear
perturbations of SFDM models.

In Sec.~\ref{sec:mass-power-spectrum}, we will analyze the suppression
of power in the mass PS and discuss the possibilities to
constrain the model with current and future galaxy surveys. We will
also discuss, from a qualitative point of view, the effect of the
aforementioned suppression in the present mass function of
halos. Finally, Sec.~\ref{sec:conclusions} will be devoted to a
general discussion of the main results and conclusions.

\section{SFDM background dynamics \label{sec:scal-field-dynam}}
For the background dynamics, we shall consider a spatially-flat
Universe  with line element $ds^2 = - dt^2 + a^2(t) \delta_{ij}
  dx^i dx^j$, and populated with standard matter fluids such as:
radiation, neutrinos, baryons, and a cosmological constant $\Lambda$,
together with dark matter described by a scalar field $\phi$ endowed
with the quadratic potential~\eqref{eq:0};  the corresponding
  Lagrangian density of the latter is $\mathcal{L}_\phi = - (1/2)
  (\partial \phi)^2 - (1/2) m^2 \phi^2$. We will refer to the model
as $\Lambda$SFDM, to emphasize the role of the scalar field as dark
matter, with a cosmological constant $\Lambda$ still playing the role
of dark energy.

\subsection{Mathematical background \label{sec:math-backgr-}}
If we use $\rho_I$ and $p_I$ to denote the energy density and
pressure, respectively, for each matter fluid in the model apart from
the scalar field, the full equations of motion of the $\Lambda$SFDM
model are:
\begin{subequations}
\label{eq:1}
  \begin{eqnarray}
    H^2 &=& \frac{\kappa^2}{3} \left( \sum_I \rho_I +
      \rho_\phi \right) \, , \label{eq:1a} \\
    \dot{H} &=& - \frac{\kappa^2}{2} \left[ \sum_I (\rho_I +
      p_I ) + (\rho_\phi + p_\phi) \right] \, , \label{eq:1b} \\
    \dot{\rho}_I &=& - 3 H (\rho_I + p_I ) \,
    , \label{eq:1c} \\
    \ddot{\phi} &=& -3 H \dot{\phi} - m^2 \phi \, , \label{eq:1d}
  \end{eqnarray}
\end{subequations}
where $\kappa^2 = 8\pi G$, a dot denotes derivative with respect to
cosmic time $t$, and $H$ is the Hubble parameter. The scalar field
energy density $\rho_\phi$ and pressure $p_\phi$ are given by the
known expressions:
\begin{equation}
  \rho_\phi = \frac{1}{2} \dot{\phi}^2 + \frac{1}{2} m^2 \phi^2 \, ,
  \quad p_\phi = \frac{1}{2} \dot{\phi}^2 - \frac{1}{2} m^2 \phi^2 \, .
\end{equation}

In order to transform the Klein-Gordon (KG) equation~\eqref{eq:1d}, we
define a new set of variables\cite{Copeland:1997et}:
\begin{equation}
  x \equiv \frac{\kappa \dot{\phi}}{\sqrt{6} H} \, , \quad y
  \equiv - \frac{\kappa m \phi}{\sqrt{6} H} \, , \quad y_1 \equiv 2
  \frac{m}{H} \, , \label{eq:2}
\end{equation}
 In writing Eqs.~\eqref{eq:2}, we are assuming that initially $\phi <
0$ so that both variables $y,y_1$ start with positive values\footnote{This
  means that the evolution of the scalar field $\phi $ starts on the left
  branch of the parabolic potential~\eqref{eq:0}, which is not a
  limitation as the system of equations is invariant under the change
  $\phi \to -\phi$. To start the evolution on the right branch, the
  signs of $y,y_1$ just need to be changed.}. As a result, the KG
equation can be written as a system of first order differential
equations in the form:
  \begin{equation}
    x^\prime = -\left(3 + \frac{\dot{H}}{H^2} \right) x + \frac{1}{2} y y_1 \,
    , \quad %\label{eq:3a} \\
    y^\prime = - \frac{\dot{H}}{H^2} y - \frac{1}{2} x y_1 \,
                 , \quad %\label{eq:3b} \\
    y^\prime_1 = - \frac{\dot{H}}{H^2} y_1 \, , \label{eq:3}
  \end{equation}
where a prime denotes derivative with respect to the number of
$e$-foldings $N \equiv \ln (a/a_i)$, with ($a_i$)  $a$ the (initial)
scale factor of the Universe. (Hereafter, we label initial conditions
with a sub-index $i$.)

We now apply a \emph{polar} change of variables to the kinetic and
potential variables in the form: $x =\Omega^{1/2}_\phi \sin(\theta/2)$
and $y = \Omega^{1/2}_\phi \cos(\theta/2)$, where $\Omega_\phi \equiv
\kappa^2 \rho_\phi/3H^2$ is the density parameter of the scalar
field. This type of transformations was first proposed
in\cite{Reyes:2010zzb} and recently applied for inflationary models in\cite{Urena-Lopez:2015odd}, but see
also\cite{Rendall:2006cq,Alho:2014fha,Jesus:2015jfa} for other similar
definitions. Firstly, we rewrite the Friedman constraint~\eqref{eq:1a}
and the acceleration equation~\eqref{eq:1b}, respectively, as
\begin{subequations}
\label{eq:5}
  \begin{equation}
    1 = \sum_j \Omega_j + \Omega_\phi \, , \quad \frac{\dot{H}}{H^2} =
    -\frac{3}{2} \left( 1 + w_{tot} \right) \, , \label{eq:5a}
  \end{equation}
where
  \begin{equation}
    \Omega_I \equiv  \frac{\kappa^2 \rho_I}{3 H^2} \, , \quad w_{tot}
    \equiv \frac{p_{tot}}{\rho_{tot}} = \sum_I \Omega_I  w_I +
    \Omega_\phi w_\phi \, , \label{eq:5b}
  \end{equation}
\end{subequations}
 and $w_I = p_I/\rho_I$ is the barotropic equation of state (EoS) of
the $I$-th matter fluid. Notice that we have defined the total EoS
$w_{tot}$ for the background dynamics in terms of the ratio of the
total pressure $p_{tot}$ to the total energy density $\rho_{tot}$,
whereas that of the scalar field is simply given by
\begin{equation}
  \label{eq:6}
  w_\phi \equiv \frac{p_\phi}{\rho_\phi} = \frac{x^2 - y^2}{x^2 +
    y^2} = - \cos \theta \, .
\end{equation}
That is, the new angular variable $\theta$ is directly related to the
scalar field EoS.

After some straightforward algebra, the KG equation~\eqref{eq:1d}
becomes:
\begin{subequations}
\label{eq:4}
  \begin{eqnarray}
  \theta^\prime &=& -3 \sin \theta + y_1 \, , \label{eq:4a} \\
  y^\prime_1 &=& \frac{3}{2}\left( 1 + w_{tot} \right) y_1 \,
  , \label{eq:4b} \\
  \Omega^\prime_\phi &=& 3 (w_{tot} - w_\phi)
  \Omega_\phi \label{eq:4c} \, .
\end{eqnarray}
\end{subequations}
Eqs.~\eqref{eq:4} are a compact and neat representation of the
KG equation, and they reveal that the true variables driving
the scalar field dynamics are $(\theta, y_1, \Omega_\phi)$, with the
field values $\phi$ and $\dot{\phi}$ now playing a secondary
role. Eqs.~\eqref{eq:4} also show that the dynamics of the quadratic
potential does not have any free physical parameters, as the
successive transformations applied to the KG equation made the field
mass $m$ become an implicit parameter in the final equations of
motion\footnote{ Strictly speaking, there is no need to consider
  $y_1$ as a separate variable, nor to take Eq.~\eqref{eq:4b} as an
  independent equation of motion. As shown in Eqs.~\eqref{eq:2}, $y_1$
is just proportional to $H^{-1}$, and then only Eqs.~\eqref{eq:4a}
and~\eqref{eq:4c} are enough to represent the KG
equation~\eqref{eq:1d} when the scalar field potential is the
quadratic one in Eq.~\eqref{eq:0}. However, $y_1$ plays an independent
role in cases with more involved scalar field potentials, and then we
will consider it as such for reference in future works (see for
instance\cite{Urena-Lopez:2015odd} in the case of inflationary
models). As we shall show, this approach will also be useful to find approximate solutions in Sec.~\ref{sec:early-late-time}
below.}.

\subsection{Early, intermediate,  and late time
  behaviors \label{sec:early-late-time}}
It is convenient to study the behavior of the scalar field under the
new equations of motion.~\eqref{eq:4}. For that we consider that there
is a special time for the beginning of the scalar field oscillations
around the minimum of the potential~\eqref{eq:0}, that we will denote by its
corresponding scale factor $a_{osc}$. Equivalently, $a_{osc}$ also
corresponds to $\theta_{osc} = \pi/2$, which is the point at which the
SFDM EoS first passes through zero, $w_\phi(\theta_{osc})
=0$\footnote{ Usual treatments define $a_{osc}$ from the approximate
  relation $m \simeq 3H(a_{osc})$. Notice that our approach gives a more
  accurate definition of $a_{osc}$ as it directly involves the scalar
  field EoS. As we shall see in Sec.~\ref{sec:initial-conditions-},
  the exact value of $a_{osc}$, which can be obtained from
  Eq.~\eqref{eq:15}, is not necessary for the numerical calculations}.

As the field in our study will be part of the matter budget, we will
also consider that the oscillations start before the time of
radiation-matter equality at $a_{eq}$, and then we put the constraint $a_{osc} <
a_{eq}$\footnote{ This is an important assumption for the
  analytical expressions in this section. However, the same
  equations of motion can also be solved approximately for the case in
  which $a_{osc} > a_{eq}$. One simple approach to do so is to find approximate
solutions Of Eqs.~\eqref{eq:4} for $a < a_{eq}$ assuming radiation
domination, and for $a > a_{eq}$ assuming matter domination. The two
solutions would be matched together at $a=a_{osc}$. This procedure
would provide good enough expressions for purposes beyond those of the
present paper.}. Thus,
for early times we will refer to the epoch of radiation domination
(RD) before the onset of the scalar field oscillations, so that $a \ll
a_{osc}$. Intermediate times will refer to $a < a_{osc}$, still within
the RD, whereas late times will strictly refer to $a > a_{osc}$. In
terms of the angular variable $\theta$, the above times can be
rephrased as: $0 < \theta \ll \pi/2$ for early times, $\theta < \pi/2$
for intermediate times, and $\theta > \pi/2$ for late times.  This
  classification is in agreement with our assumption that at early
  times $\phi < 0$, so that the scalar field starts its journey on the
  left branch of the quadratic potential~\eqref{eq:0}. Moreover, that
  $\theta >0$ also implies, via Eqs.~\eqref{eq:2} and the polar change
  of variables, that initially $\dot{\phi} >0$, so that the field
  rolls down towards $\phi=0$.

Before proceeding further, we note that in our scheme the SFDM mass
can be determined from the initial condition of the second potential
variable as: $m = y_{1i} H_i/2$, see also Eqs.~\eqref{eq:2}. In
general terms, we can expect that at early enough times $m/H_i \ll 1$,
which directly indicates also that $y_{1i} \ll 1$. In terms of the
original KG equation~\eqref{eq:1d}, this implies that the scalar field
dynamics is dominated by the friction term containing the Hubble
parameter $H$, which then puts the field into a slow-roll phase
with its energy density dominated by the potential term; this in turn
translates into $x \ll y$, and then for the angular variable:
$\theta_i \ll 1$. A similar result is expected for $\Omega_{\phi i}
\ll 1$ if SFDM is to be subdominant during RD.

The above simple estimations indicate that the scalar field variables
must have initial values very close to the following critical point of
the dynamical system~\eqref{eq:4}: $(\theta=0,y_1
=0,\Omega_\phi=0)$. Under this assumption, Eqs.~\eqref{eq:4} can be
linearised around this critical point as:
\begin{equation}
  \theta^\prime \simeq -3 \theta + y_1 \, , \quad y^\prime_1 \simeq 2
  y_1 \, , \quad \Omega^\prime_\phi \simeq 4 \Omega_\phi \,
  , \label{eq:7}
\end{equation}
where we have considered that $w_{tot} = 1/3$, as expected in RD, and
also that $w_\phi \simeq -1$ and $\sin \theta \simeq \theta$ for
$\theta \ll 1$. The linear system~\eqref{eq:7} has the following
solutions in terms of the scale factor $a$:
\begin{subequations}
  \label{eq:8}
  \begin{eqnarray}
    \theta &=& (1/5) y_1 + C (a/a_i)^{-3} \, , \label{eq:8a} \\
    y_1 &=& y_{1i} (a/a_i)^2 \, , \label{eq:8b} \\ 
    \Omega_\phi &=& \Omega_{\phi i} (a/a_i)^4 \, , \label{eq:8c}
  \end{eqnarray}
\end{subequations}
with $C$ an integration constant. The angular variable $\theta$ has
growing and decaying modes, and when the latter disappears we find
that $5 \theta \simeq y_1$. We shall call this relation the
\emph{attractor solution} at early times between variables $\theta$
and $y_1$. As for the scalar field density parameter, its
solution~\eqref{eq:8c} indicates that the SFDM energy density is
constant during RD: $\rho_\phi \simeq \Omega_\phi \rho_r = {\rm
  const}$\footnote{As a side result, one can see that the
  attractor solutions implies that the evolution of the scalar field
  EoS at early times is approximately given by $w_\phi \simeq -1 +
  \theta^2/2 = -1 + (2/25)m^2/H^2$, which coincides at the lowest
  order with Eq.~(A31) in\cite{Hlozek:2014lca}}.

For intermediate times, a more general solution of Eqs.~\eqref{eq:4a}
and~\eqref{eq:4b} can be found perturbatively, as $\theta$ and $y_1$
remains reasonably small up to the time before the onset of
oscillations at $\theta = \pi/2$. Moreover, $\theta$ has a monotonic
behavior, and then we can propose the ansatz:
\begin{equation}
  y_1 (\theta) = \sum_{j=1} k_j \theta^j \, , \quad k_j =
  \mathrm{const}. \label{eq:9}
\end{equation}
We would like to emphasize that $y_1(\theta)$ is all that is needed to
find a full solution of the scalar field
equations~\eqref{eq:1}. Indeed, once with the values of $k_j$ at
hand we can write Eq.~\eqref{eq:4a} as:
\begin{equation}
  \theta^\prime = (k_1 -3)\theta + k_2 \theta^2 + (k_3 + 1/2)
  \theta^3 + \ldots \, . \label{eq:11}
\end{equation}

Eq.~\eqref{eq:11} can in principle be integrated at any desired order
in $\theta$. For our purposes, it is enough to consider in
Eq.~\eqref{eq:6} an expansion up to the third order, and then
$y_1(\theta) = k_1 \theta + k_2 \theta^2 + k_3 \theta^3$. The
key step to find the coefficients $k_j$ is to write
Eq.~\eqref{eq:4b} in the form
\begin{equation}
  \label{eq:14}
  y^\prime_1 = \frac{dy_1}{d\theta} \theta^\prime =
  \frac{dy_1}{d\theta} (-3\sin \theta + y_1) = 2y_1 \, ,
\end{equation}
where we again assummed RD with $w_{tot} = 1/3$ for the last term on
the rhs. After expanding both sides of Eq.~\eqref{eq:14} in powers of
$\theta$, the resultant polynomials must have the same coefficients,
and from this condition we obtain $k_1=5$, $k_2 =0$, and
$k_3 = -5/18$. Finally, Eq.~\eqref{eq:11} can be integrated to find:
\begin{equation}
  \frac{\theta^2}{\theta^2_i} \left( \frac{9 +
      \theta^2_i}{9 + \theta^2} \right) = (a/a_i)^{4} \,
  , \label{eq:15}
\end{equation}
which could be easily inverted to find $\theta = \theta(a)$ for $a <
a_{osc}$. If necessary, we can also calculate\footnote{ Notice that
  according to this solution the ratio of the SFDM mass to the Hubble
  parameter at the start of the oscillations is $m/H_{osc}
  \simeq 3.39$.} $y_1 = 5 \theta - 5\theta^3/18$, whereas
$\Omega_\phi$ would still be given by Eq.~\eqref{eq:8c}.

At late times $a> a_{osc}$, the scalar field field starts to oscillate
rapidly around the minimum of the potential with a decreasing
amplitude, and then we expect $y_1$ to grow very quickly and to become
the dominant term in Eq.~\eqref{eq:4a}: $\theta^\prime \simeq
y_1$. Changing back the derivative in the latter equation to the
cosmic time $t$, we find that:
\begin{equation}
  \dot{\theta} \simeq 2m  \to \theta \simeq 2m t \, , \quad {\rm and}
  \quad w_\phi \simeq - \cos(2m t) \, . \label{eq:16}
\end{equation}
Irrespective of the domination era after the onset of oscillations, we
can see that the SFDM EoS will oscillate around zero with a fixed
frequency, in terms of the cosmic time, given by the field mass
$m$. The oscillations happen in a time scale much smaller than that of
the expansion rate of the Universe, which means that $H^{-1} >
m^{-1}$. The Hubble parameter is a decreasing function of time, and we
can only expect that $y_1 = 2m/H \gg 1$ as the Universe continues its
expansion for $a > a_{osc}$.

For this reason, it is generically considered that the SFDM energy
density evolves, on average, with a matter-like EoS: $\langle w_\phi
\rangle \simeq 0$\cite{Turner:1983he,Ratra:1990me}. This expectation
has also been confirmed by accurate numerical
simulations\cite{Sahni:1999qe,Matos:2000ng,Matos:2000ss,Magana:2012xe}
(see also Fig.~\ref{fig:2} in the Appendix~\ref{sec:cut-proc-rapidly}),
and then we can safely take for granted that the SFDM background
density will evolve as a matter component, $\rho_\phi \propto a^{-3}$,
irrespectively of the presence of other fluid components.

\subsection{Initial conditions \label{sec:initial-conditions-}}
In order to find numerical solutions of the dynamical
system~\eqref{eq:4}, all that remains is to find appropriate initial
conditions $(\theta_i,y_{1i},\Omega_{\phi i})$, so that we can
get the desired SFDM contribution $\Omega_{\phi 0}$ at the present
time. This is done in most CMB codes, in terms of the standard field
scheme, by the implementation of a shooting procedure to determine the
correct initial values $\phi_i$ and
$\dot{\phi}_i$\cite{Lesgourgues:2011re}. However, it is still
necessary to find a guess value, commonly obtained trough a cumbersome
calculation\cite{Marsh:2010wq,Hlozek:2014lca}, to initiate
the shooting method. As we shall show below, there is a more direct
path to find the correct initial conditions in terms of our new
dynamical variables.

We start by writing down an equation for the initial value of the
SFDM density parameter $\Omega_{\phi i}$. For that, we first recall
that once the scalar field starts to oscillate rapidly, its energy
density redshifts as $a^{-3}$, so that its ratio with respect to the
radiation component is: $\Omega_\phi / \Omega_r = (\Omega_{\phi 0} /
\Omega_{r0}) a$  for $a > a_{osc}$. In particular, we can write
$\Omega_{\phi \, osc} \simeq a_{osc} \Omega_{r osc} (\Omega_{\phi 0} /
\Omega_{r0}) \simeq a_{osc} (\Omega_{\phi 0} / \Omega_{r0})$, where
the last expression takes into account that $\Omega_{r osc} = 1$ if
the Universe is still dominated by radiation at $a= a_{osc}$. From
Eq.~\eqref{eq:8c}, we also know that $\Omega_{\phi \, osc} \simeq
(a_{osc}/a_i)^4 \Omega_{\phi i}$, and all together with
Eq.~\eqref{eq:15} we finally find that:
\begin{equation}
  \Omega_{\phi i} \simeq a_i \frac{\Omega_{\phi 0}}{\Omega_{r0}} \left(
    \frac{a_i}{a_{osc}} \right)^3 = a_i \frac{\Omega_{\phi
      0}}{\Omega_{r0}} \left[ \frac{4\theta^2_i}{\pi^2} \left(
      \frac{9 + \pi^2/4}{9 + \theta^2_i} \right) \right]^{3/4}
  \, . \label{eq:19}
\end{equation}

The search for good initial conditions is now quite simple. We first
need to set up by hand the desired contributions of the radiation and
the SFDM components at the present time: $\Omega_{r0}$, $\Omega_{\phi
  0}$, together with the initial value of the scale factor
$a_i$. Then, we choose an initial value $\theta_i$, use the attractor
solution of Eq.~\eqref{eq:8a} at early times to find: $y_{1i} =
5\theta_i$, and finally take Eq.~\eqref{eq:19} to find the
corresponding initial value $\Omega_{\phi i}$.

To conclude this section, it must be said that the ratio of the SFDM
mass to the initial Hubble parameter is a \emph{derived} value
calculated from the initial conditions: in terms of the attractor
relation for the potential variable $y_1$, we find that $m/H_i =
y_{1i}/2 = (5/2) \theta_i$. A more useful formula can be written in
terms of the present Hubble parameter $H_0$ as:
\begin{eqnarray}
    \frac{m}{H_0} &=& \frac{H_i}{H_0} \frac{y_{1i}}{2} = \frac{5}{2}
   \Omega^{1/2}_{r0} a^{-2}_i \theta_i \, , \label{eq:10}
\end{eqnarray}
where we have again assumed RD in the form $H^2_i = (\kappa^2/3)
\rho_{r0}a^{-4}_i$ to write the last expression on the
rhs. Eq.~\eqref{eq:10} shows again the interplay of the initial
conditions to set the value of the scalar field mass: everything else
the same, different masses $m$ are just obtained by taking different
values of the initial condition $\theta_i$.

\subsection{Modifications in CLASS}
\label{sec:modifications-class}
The scheme described above to evolve the scalar field equations of
motion was incorporated in an amended version of the freely-available code
CLASS\cite{Lesgourgues:2011re,Blas:2011rf,Lesgourgues:2011rg}\footnote{It
  must be noticed that the equations of motion~\eqref{eq:4} are
  written with derivatives with respect to the number of $e$-foldings
  $N$. The derivatives can be changed to conformal time $\tau$ (the
  preferred time in CMB codes) via the well-known transformation
  $(d/d\tau) = a H(d/dN) = \mathcal{H}(d/dN)$, where $\mathcal{H}$ is
  the Hubble parameter itself given in terms of the conformal
  time}. In order to be consisten with the background evolution of
other matter components within CLASS, the scalar field energy density and
pressure were calculated, respectively, from the expressions:
 \begin{equation}
   \rho_\phi = \frac{\Omega_\phi}{1- \Omega_\phi} (\rho_r + \rho_m +
   \rho_\Lambda) \, , \quad p_\phi = -\cos \theta \cdot \rho_\phi \,
   . \label{eq:17}
 \end{equation}
The above formulae are found under the assumption that $\Omega_\phi =
\rho_\phi/ \rho_{tot}$, where $\rho_{tot}$ is the total energy density
at any given time. This assumption is in agreement with a spatially
flat Universe, for which the total energy density coincides with the
critical one $\rho_{tot} = \rho_{crit} = 3H^2/8\pi G$. Notice
  that our scheme does not exclude the presence of other dark matter
  components (like CDM) apart from SFDM. Actually, following the given
  structure within CLASS, all actual matter contributions are put in
  by hand, except for that of SFDM, which is calculated the last from
  the fulfillment of the Friedman constraint for a spatially flat
  Universe (which is also the scheme within CLASS to deal with quintessence
  fields). Then, the amended code can also be used to study mixed dark
  matter models (as in\cite{Hlozek:2014lca}), and no change for such a
  case would required in the formalism presented throughout this
  work. However, for our present purposes, we shall consider that SFDM
  is the only dark matter component.

Even though Eqs.~\eqref{eq:4} are more convenient for accurate
numerical solutions than the traditional field approach, they still
require a highly efficient solver of differential equations to deal
properly with the numerical stiffness of the scalar field
oscillations. In order to circumvent this difficulty, we implemented a
cut-off procedure on the trigonometric functions in the form:
\begin{equation}
  \{\cos_\star \gamma, \sin_\star \gamma \} \equiv (1/2) \left[ 1 -
    \tanh(\gamma^2 - \gamma^2_\star) \right] \{ \cos \gamma, \sin
  \gamma \} \, , \label{eq:18}
\end{equation}
and then $ \{\cos_\star \gamma, \sin_\star \gamma \} \to 0$ for
$\gamma > \gamma_\star$. The threshold value $\gamma_\star$ was chosen
empirically to neglect the trigonometric functions everywhere once the
equation solver in CLASS was no longer able to keep an accurate track
of the rapid oscillations of the scalar field. After some
experimentation, we found that a convenient value is $\gamma_\star =
10^2$.

An illustrative physical example is the scalar field EoS, which was
written inside CLASS as: $w_\phi = - \cos_\star \theta$. As we saw
in Sec.~\ref{sec:early-late-time}, during the phase of rapid
oscillations $\theta \simeq y_1/2 \simeq m/H$, and then in this case the
threshold value $\gamma_\star$ means $\theta_\star \simeq m/H_\star
\simeq 10^2$. In other words, $w_\phi$ is set to zero once the time
scale for the expansion of the Universe is two orders of magnitude
larger than the period of one oscillation in the EoS,  irrespective of
the value of the considered SFDM mass $m$. As explained in
more detail in the Appendix~\ref{sec:cut-proc-rapidly}, the
implemented numerical cut-off~\eqref{eq:18} worked well, and did not
introduce spurious effects into the numerical solutions.

As for the initial conditions, it must be said that Eq.~\eqref{eq:19}
is not quite the correct initial condition for the SFDM density
parameter, and we still needed to finely tune $\Omega_{\phi i}$ at the
beginning of every numerical run. For that, we wrote $\Omega_{\phi i}
= A \times$Eq.~\eqref{eq:19}, where the value of $A$ was adjusted with
the shooting method already implemented within CLASS for scalar field
models. A few iterations of the shooting routine were always enough to
find the correct $\Omega_{\phi i}$ that leads to the desired
$\Omega_{\phi 0}$ with a very high precision; in all instances it was
found that $A = \mathcal{O}(1)$.

\begin{figure*}[htp!]
\centering
\includegraphics[width=0.49\textwidth]{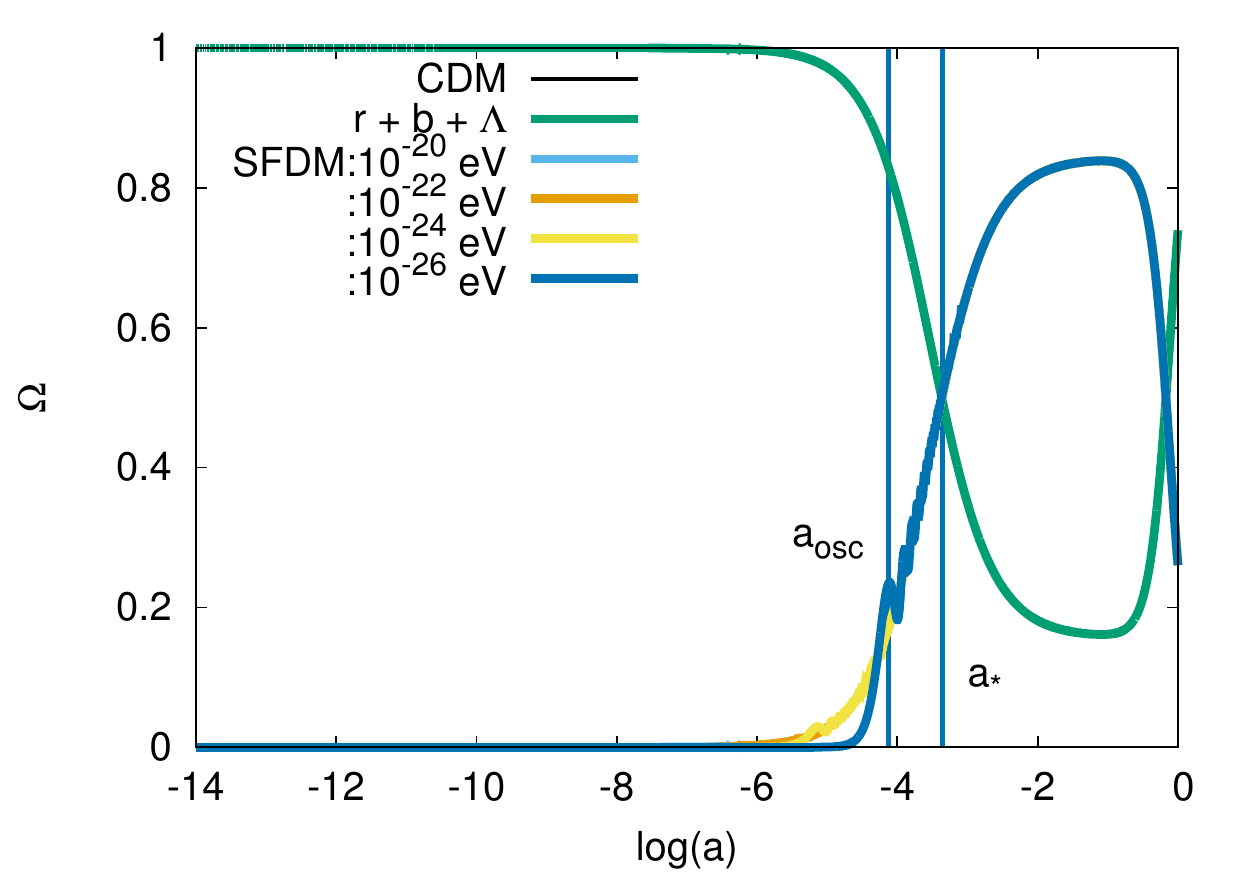}
\includegraphics[width=0.49\textwidth]{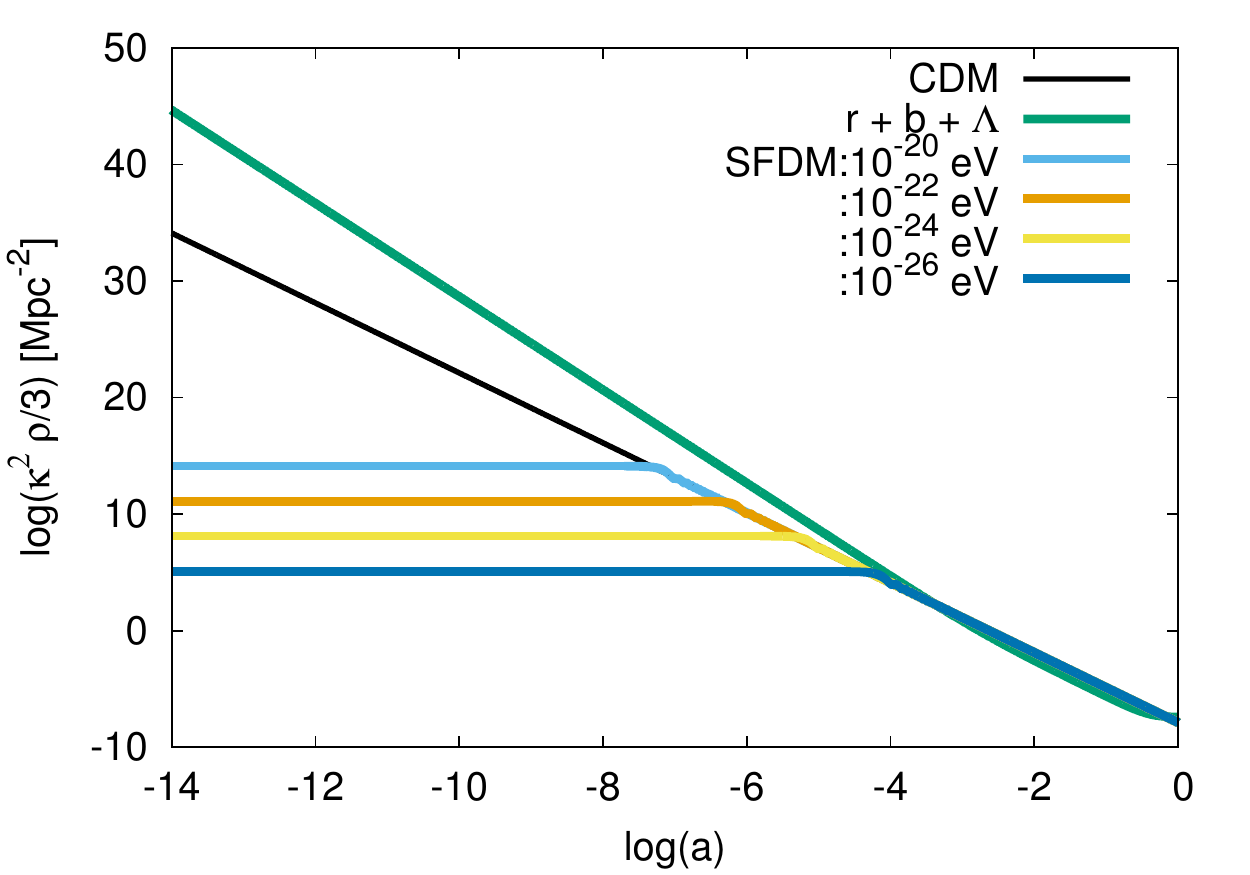}
\caption{\label{fig:1} Numerical solutions of the background
  equations~\eqref{eq:1} and~\eqref{eq:4} with a scalar field as dark
  matter at late times. The label CDM in the plots refers to the
  standard dark matter component, whereas $r$, $b$ and $\Lambda$ stand
  for radiation (photons and neutrinos), baryons, and a cosmological
  constant. Shown are: (Left) the density parameters $\Omega$ (see
  Eqs.~\eqref{eq:5}), and (Rigth) the energy densities $\rho$, of the
  different matter components. Notice that the standard evolution of
  the $\Lambda$CDM model is fully recovered at late times. The onset
  of the field oscillations is manifest in the wiggles present in the
  curves, although the most noticeable cases correspond to the
  lightest masses, for which the oscillations start closely to the
  time of radiation-matter equality at $a \simeq 10^{-4}$.  As an
  example, the vertical lines indicate the start of the oscillations
  at $a_{osc}$, and the application of the cut-off for rapid
  oscillations at $a_\ast$, for the lightest case of $m= 10^{-26} {\rm
    eV}$. See
  Secs.~\ref{sec:early-late-time},~\ref{sec:modifications-class}, and
  the Appendix~\ref{sec:background-variables}, for more details.}
\end{figure*}

\subsection{Numerical results for the
  background \label{sec:numer-results-backgr}}
Some numerical results from CLASS of background quantities are shown
in Figs.~\ref{fig:1}, for a Universe in which the scalar field is the
only dark matter component, that is, $\Omega_{\phi0} = \Omega_{CDM
  0}$, whereas all other matter components (radiation, baryons, and
$\Lambda$) are the same as reported by the Planck collaboration for
the concordance model $\Lambda$CDM\cite{Martin:2015dha}. The mass of
the SFDM particle in the given examples is indicated in the plots, and
in each case was calculated from the initial condition $\theta_i$
through Eq.~\eqref{eq:10}.

A general feature in all the numerical examples is that the SFDM
energy density $\rho_\phi$ is constant at very early times, when its
dynamics is dominated by its potential energy. Once the oscillations
start, the energy density scales as a matter component, and the
evolution of the density parameters clearly show that the subsequent
epochs of matter and $\Lambda$ domination are indistinguishable from
those of $\Lambda$CDM. The onset of oscillations can be clearly seen
in the evolution of the density parameter $\Omega_\phi$, specially in the
cases with the lightest masses. Actually, the oscillations start
well before the time of radiation-matter equality except for the
lightest case $m \simeq 10^{-26} {\rm eV}$.  A detailed
  description of the solution of $\Omega_\phi$ can be found in the
  Appendix~\ref{sec:cut-proc-rapidly}, where we also show that the
  cut-off~\eqref{eq:18} helps CLASS to calculate effortlessly the
  right solution at late times.

\section{SFDM linear perturbations \label{sec:line-axion-pert}}
The linear perturbations of scalar fields, and its influence in the
CMB and other related observables, have been studied intensively in the specialized literature under a variety of
different circumstances, see for instance\cite{Ferreira:1997au,Ferreira:1997hj,Perrotta:1998vf,Matos:2000ss,Magana:2012xe,Park:2012ru,Noh:2013osa,Hwang:2009js,Marsh:2015daa,Marsh:2010wq,Hlozek:2014lca} and
references therein. Here, we aim to present a new study of field
perturbations that, in many aspects, mirrors the one we presented for
the background in Sec.~\ref{sec:scal-field-dynam}.

Following common wisdom, here we shall assume the synchronous gauge of
metric perturbations, with the line element $ds^2 = -dt^2 + a^2(t)
(\delta_{ij} + h_{ij})dx^i dx^j$. Notice that we will not work with
the standard conformal time $\tau$, and then a dot still means
derivative with respect to the cosmic time $t$. The scalar field is
given by $\phi(x,t) = \phi(t) + \varphi(x,t)$, where $\phi(t)$ is the
background (homogeneous) field, and $\varphi$ is its linear perturbation. Thus, the linearized KG equation for a Fourier mode
$\varphi(k,t)$
reads\cite{Ratra:1990me,Ferreira:1997au,Ferreira:1997hj,Perrotta:1998vf}:
\begin{equation}
  \ddot{\varphi} = - 3H \dot{\varphi} - (k^2/a^2 + m^2) \varphi -
  \frac{1}{2} \dot{\phi} \dot{h} \, , \label{eq:13}
\end{equation}
where $h$ is the trace of scalar metric perturbations (with $\dot{h}$
known as the metric continuity), and $k$ is a comoving wavenumber. The
perturbations in density $\delta \rho_\phi$, pressure $\delta p_\phi$,
and velocity divergence $\Theta_\phi$, are given, respectively, by\cite{Perrotta:1998vf,Hu:1998kj,Ferreira:1997hj}:
\begin{subequations}
\label{eq:21}
  \begin{eqnarray}
    \delta \rho_\phi &=& \dot{\phi}  \dot{\varphi} + \partial_\phi V \,
    \varphi \, , \label{eq:21a} \\
    \delta p_\phi &=& \dot{\phi}  \dot{\varphi} - \partial_\phi V \,
    \varphi \, , \label{eq:21b} \\
    (\rho_\phi + p_\phi) \Theta_\phi &=& (k^2/a) \dot{\phi} \varphi \, . \label{eq:21c}
  \end{eqnarray}
\end{subequations}

The standard field approach for scalar field perturbations is to solve
Eq.~\eqref{eq:13} for a given background evolution to find $\varphi$
and $\dot{\varphi}$, which are then used to calculate the perturbative
quantities~\eqref{eq:21}. At a heuristic level, and for a rapidly
oscillating scalar field, we can see that its linear perturbations
grow if the fields in the KG equations~\eqref{eq:1c} and~\eqref{eq:13}
have similar natural frequencies, which would happen for scales such
that $k \ll m$, whereas in the opposite case $k \gg m$, we expect a suppression of the perturbations.

The true behavior of scalar field perturbations is actually more
complex, and numerical solutions are unavoidable. However,
Eq.~\eqref{eq:13} suffers from the same stiffness problem of the
background equations, mainly because of the rapidly oscillatory terms
that arise on the rhs from the presence of $m^2$ and $\dot{\phi}$,
which also renders it impractical for numerical use in CMB
codes\cite{Matos:2000ss,Marsh:2010wq,Hlozek:2014lca}. An alternative,
and second approach to the study of scalar field perturbations, is to
transform Eq.~\eqref{eq:13} into an equivalent set of fluid equations,
as was first done in the so-called formalism of generalized dark matter\cite{Hu:1998kj}. Such transformation is still impractical for
numerical purposes (as already noted in\cite{Hu:1998kj}) in the case
of rapidly oscillating scalar fields, mostly because some fluid terms,
like the sound speed $c^2_s = \delta p_\phi/ \delta \rho_\phi$, are
ill behaved during the oscillating phase of the scalar field.

Such difficulties in the fluid approximation have not deterred its
ample use in the specialized
literature\cite{Amendola:2005ad,Park:2012ru,Noh:2013osa,Hwang:2009js,Marsh:2015daa,Marsh:2010wq,Hlozek:2014lca}. One
common corrective step in this method, that was firstly suggested
in\cite{Ratra:1990me}, is to assume an approximate ansatz for the
field oscillations, and then to derive from it \emph{averaged} fluid
equations freed from any troublesome
behavior\cite{Hwang:2009js,Park:2012ru,Hlozek:2014lca}. Although
highly popular, this approach requires some intrusive pre-handling of
the equations of motion that obscures the physical mechanisms at work
in the evolution of SFDM perturbations.

We propose a third approach, based in a new field transformation of Eq.~\eqref{eq:13} quite similar to that we used for the background
variables:
\begin{subequations}
  \label{eq:22}
  \begin{eqnarray}
    u &=& \sqrt{\frac{2}{3}} \frac{\kappa \dot{\varphi}}{H} = 
    -\Omega^{1/2}_\phi e^\alpha \cos(\vartheta/2) \, , \label{eq:22a} \\ 
    v &=& \sqrt{\frac{2}{3}} \frac{\kappa m \varphi}{H} =
    -\Omega^{1/2}_\phi e^\alpha \sin(\vartheta/2) \, . \label{eq:22b}
  \end{eqnarray}
\end{subequations}
For a simple interpretation of the new variables $\vartheta$ and
$\alpha$, it is useful to define the density $\delta_\phi \equiv
\delta \rho_\phi/ \rho_\phi$, and pressure $\delta_{p_\phi} \equiv \delta
p_\phi/ \rho_\phi$ contrasts, respectively, from the physical
quantities~\eqref{eq:21}. In summary, we find:
 \begin{subequations}
   \label{eq:26}
  \begin{eqnarray}
    \delta_\phi &=& - e^\alpha \sin(\theta/2 - \vartheta/2) \,
    , \label{eq:26a} \\
    \delta_{p_\phi} &=& - e^\alpha \sin(\theta/2 + \vartheta/2) \,
    , \label{eq:26b} \\
    (\rho_\phi + p_\phi) \Theta_\phi &=& -\frac{k^2}{am} \rho_\phi
                                         e^\alpha \sin(\theta/2)
                                         \sin(\vartheta/2) \,
                                         . \label{eq:26c}
  \end{eqnarray}
\end{subequations} 

The appearance of oscillations in the perturbative quantities can be
easily spotted in Eqs.~\eqref{eq:26}; they result from the combination
of the background and perturbative angular variables $\theta$ and
$\vartheta$. Moreover, we can see that $e^\alpha$ can be simply
interpreted as the amplitude of the scalar field density and pressure
contrasts.

After long, but otherwise straightforward, calculations, we find that
the equations of motion of the new variables, as derived from the
perturbed KG equation~\eqref{eq:13}, are:
\begin{subequations}
  \label{eq:25}
  \begin{eqnarray}
    \vartheta^\prime &=& 3 \sin \vartheta + y_1 + 2 \omega \sin^2
    (\vartheta/2) - 2 e^{-\alpha} h^\prime \sin(\theta/2) \sin(\vartheta/2) \,
    , \label{eq:25a} \\
    \alpha^\prime &=&  - \frac{3}{2} \left( \cos \vartheta + \cos \theta
                      \right) - \frac{\omega}{2} \sin \vartheta +
                      e^{-\alpha} h^\prime \sin(\theta/2)
                      \cos(\vartheta/2) \, . \label{eq:25b}
  \end{eqnarray}
\end{subequations}
There are two remarkable features in Eqs.~\eqref{eq:25}. The first
one is that the Hubble parameter $H$ does not appear explicitly, and
then the only connection of the perturbation variables to the
background expansion is through the presence of $\theta$ and $y_1$. The
second feature is that $y_1$, in a similar way as in
Eq.~\eqref{eq:4a}, should at some point become the dominant term in
Eq.~\eqref{eq:25a}, and then we must expect that at late times
$\vartheta \simeq \theta = 2mt$. Hence, the perturbation variables $\varphi$
and $\dot{\varphi}$, see Eqs.~\eqref{eq:22}, will also be rapidly
oscillating functions just as their background counterparts.

Interestingly enough, Eqs.~\eqref{eq:25} quite naturally indicate the
presence of a characteristic wavenumber in scalar field
perturbations. For simplicity, we have denoted the ratio $\omega =
k^2/k^2_J$, where
\begin{subequations}
\label{eq:23}
\begin{equation}
  k_J \equiv a \sqrt{mH} \, , \label{eq:23a}
\end{equation}
 Notice that $k_J$ is a time-dependent quantity that has been
 previously identified, sometimes just heuristically, as the comoving
 Jeans wavenumber of linear perturbations in scalar field models,
 see\cite{Hu:1998kj,Hu:2000ke,Matos:2000ng,Woo:2008nn,Marsh:2010wq,Hlozek:2014lca}
 and references therein. Its present, and then physical, value is:
\begin{equation}
  \frac{k_{J0}}{ {\rm Mpc}^{-1}} = 0.68 \left( \frac{m}{10^{-23} {\rm
        eV}} \right)^{1/2} \left( \frac{H_0}{\rm km \, s^{-1} \,
      Mpc^{-1}} \right)^{1/2} \, . \label{eq:23b}
\end{equation}
\end{subequations}
Actually, if the initial conditions are given in RD, we find that the
variation of the Jeans wavenumber is just proportional to the present
density parameter of radiation: $k_{Ji}/k_{J0} = a_i \sqrt{H_i/H_0} =
\Omega^{1/4}_{r0}$. This also means that  in a typical cosmological
evolution $k^2/k^2_J \lesssim 1$ for most of the scales of interest in
structure formation, except in the case of extremely light masses $m
\ll 10^{-23} {\rm eV}$. A complete study of the effects of the Jeans
wave number $k_J$ in the evolution of SFDM perturbations will be
presented in Sec.~\ref{sec:suppression-power-at} below.

\subsection{Early time behavior \label{sec:early-time-behavior}}
As for the background case, it is convenient to make a description of
the early time behavior of the scalar field perturbations from the
point of view that Eqs.~\eqref{eq:25} is a dynamical system, with at
least one meaningful critical point as $a \to 0$ ($N \to -\infty$),
which will then act as the source point for the physical solutions. In
contrast to the background equations~\eqref{eq:4}, the dynamical
system~\eqref{eq:25} is not autonomous due to the presence of the
background quantities $\theta$ and $y_1$, and the metric term
$h^\prime$, which together act as an external time-dependent force
term for the perturbations.

If we assume RD, then we know from Eqs.~\eqref{eq:8} that the behavior
of the background variables in terms of the scale factor is: $5 \theta
= y_1 = y_i (a/a_i)^2$, whereas that of the metric perturbation must
be $h = h_i(a/a_i)^2$, where $h_i$ is a constant. This last expression
comes from the well known solutions of linear perturbations in a RD
universe with adiabatic initial conditions,
see for instance\cite{Perrotta:1998vf,Hlozek:2014lca} for more
details. In order to find their critical point, we rewrite
Eqs.~\eqref{eq:25} in the form:
\begin{subequations}
  \label{eq:39}
  \begin{eqnarray}
    \vartheta^\prime &=& 2 \sin(\vartheta/2) \left[ 3 \cos
                         (\vartheta/2) + \omega \sin (\vartheta/2) -
                         e^{-\alpha} h^\prime \sin(\theta/2) \right] +
                       y_1 \, , \label{eq:39a} \\
    \alpha^\prime &=& - \cos(\vartheta/2) \left[ 3 \cos
                      (\vartheta/2) + \omega \sin (\vartheta/2) -
                      e^{-\alpha} h^\prime \sin(\theta/2) \right] + 3
                    \sin^2(\theta/2) \, . \label{eq:39b}
  \end{eqnarray}
By simple inspection in Eq.~\eqref{eq:39a}, we see that one critical
point as $a \to 0$ is $\vartheta = 0$. If we substitute this value
into Eq.~\eqref{eq:39b} we find that:
\begin{equation}
  \label{eq:37}
  \alpha^\prime \simeq -3 + e^{-\alpha} h^\prime \sin(\theta/2) + 3
  \sin^2(\theta/2)\, .
\end{equation}
\end{subequations}
If we drop the last term in Eq.~\eqref{eq:37} for being of a higher
order in  the expansion of $\theta$ (which is a small quantity),
we find that a first integral is: $e^\alpha \simeq (2/7) h
\sin(\theta/2)$. Whereas $\vartheta$ reaches a true critical value,
the critical condition on $\alpha$ just indicates its asymptotic behavior close to the point $a=0$: $\alpha \sim 4N$
\footnote{In this respect, we cannot find a true critical point in the
  phase space spanned by the variables $(\vartheta, \alpha)$. However,
  the critical point exists, at least formally, if the dynamical
  variables are $(\vartheta,e^\alpha)$, and would correspond to
  $(\vartheta = 0,e^\alpha \to 0)$.}.

We can now calculate the solution of $\vartheta$ to the next order
close to the critical point. Considering a perturbation of the form
$\vartheta = 0 + \delta \vartheta$, then we find that at the lowest
order Eq.~\eqref{eq:39a} becomes: $\delta \vartheta^\prime = -4 \delta
\vartheta + y_1$, from which we readily obtain that the growing
solutions is $\delta \vartheta = (1/6) y_1 = (5/6) \theta$. Therefore,
the early-time solutions of the scalar field perturbations are
represented by the expressions:
\begin{equation}
  \label{eq:45}
  \vartheta = (5/6) \theta \, , \quad e^\alpha = (1/7) h \theta
  \, .
\end{equation}
Eqs.~\eqref{eq:45} can be considered the \emph{attractor} solutions of
SFDM perturbations in the very early universe, and from them we can
extract the right initial conditions for the scalar field
perturbations. One important property of Eqs.~\eqref{eq:45} is that
they do not depend on the wavenumber $k$, and then they are of general
applicability for all scales.

Notice also that Eqs.~\eqref{eq:45} mean that the behavior of the
scalar field density contrast close to the critical point, from Eq.~\eqref{eq:26a}, can be cast in the form
\begin{subequations}
\begin{equation}
  \label{eq:42}
  \delta_\phi \simeq - \frac{1}{21} h \sin^2(\theta/2) = - \frac{1}{42}
  h (1+w_\phi) \, ,
\end{equation}
and then the adiabatic condition $\delta_\phi=0$ is accomplished as
the scalar field EoS $w_\phi \to -1$ (see Eq.~\eqref{eq:6}) for $a
\to 0$\cite{Perrotta:1998vf,Hlozek:2014lca}. Likewise, we can
calculate the asymptotic expressions for the pressure contrast and the
velocity divergence; from Eqs.~\eqref{eq:26b} and~\eqref{eq:26c} we
find, respectively, that:
\begin{eqnarray}
  \delta_{p_\phi} & \simeq& - \frac{11}{21} h \sin^2(\theta/2) = -
                            \frac{11}{42} h (1+w_\phi) \,
                            , \label{eq:48a} \\
  (\rho_\phi + p_\phi) \Theta_\phi &\simeq& - \frac{5}{21} \frac{k^2}{am}
                                       \rho_\phi h \sin^3(\theta/2) \,
                                       . \label{eq:48b}
\end{eqnarray}
We see that both quantities also vanish as $a \to 0$.
\end{subequations}

\subsection{Gauge ambiguities and matter perturbations}
\label{sec:gauge-ambig-matt}
One of the favourite gauges for the numerical evolution of
perturbations is the synchronouns gauge, mostly for its reliability
and straightforward formulation for different matter components. It is
known, however, that this gauge can only be completely determined if
additional constraints are taken into account. The standard procedure
is to fix its remaining degree of freedom by considering that all
calculations are performed in the comoving frame of the cold dark
matter component, in which the corresponding velocity perturbation
vanishes, i.e. $\Theta_{\rm CDM}=0$\cite{Ma:1995ey}. It is clear that one
must then be concerned about this gauge fixing in the cases where a
scalar field is the dark matter, and specially if it is the only such component\cite{Christopherson:2012kw}. 

Most papers in the literature assume, under the fluid approximation of
the equations of motion, that a rest framework can be established for
SFDM much in the same way as for CDM; the physical condition that must
be satisfied is
\begin{equation}
  \label{eq:36}
  (\rho_\phi + p_\phi ) \Theta_\phi = 0 \, .
\end{equation}
This certainly can be done once the scalar field is oscillating and
redshifting as matter\cite{Hwang:2009js,Park:2012ru,Hlozek:2014lca,Marsh:2010wq}, but
the concern about the gauge issue remains if one wishes to start a
cosmological evolution well before the onset of the field
oscillations. Although it is possible to consider the Newtonian
gauge\cite{Magana:2012xe}, or better a gauge-invariant
evolution\cite{Alcubierre:2015ipa},  we adopted the practical
  point of view that all calculations are made in a frame comoving
  with a pressureless dust component. This is the standard assumption
  in CLASS, and actually we still needed to consider a tiny
  contribution of CDM ($\Omega_{\rm CDM 0} < 10^{-6}$) so that CLASS
  would be able to run satisfactorily.

 In this sense, because we are not working in the frame comoving
  with the scalar field, we should not expect the accomplishment of
  Eq.~\eqref{eq:36}, but rather, as we shall see in
  Secs.~\ref{sec:large-scales-k} and~\ref{sec:small-scales-k} below,
  the scalar field perturbations must behave as those of a pressureless
  component. Such behavior is particularly clear in the case of the
  velocity perturbations, as the quantity $(\rho_\phi + p_\phi )
  \Theta_\phi$ evolves quite similarly to that of baryons, see
  Sec.~\ref{sec:gener-solut-pert} and Fig.~\ref{fig:7}.

Finally, there is some concern also about the inclusion of SFDM
perturbations into the matter spectrum, see for instance the
discussion in Sec.~IV B in\cite{Hlozek:2014lca}. Here we will adopt a
conservative point of view, also implicitly assumed within the
perturbations module of CLASS, that the term \emph{matter
  perturbation} comprises the perturbations coming from all components
behaving as CDM at late times. This criterion is certainly fulfilled
by the SFDM models studied here, as they are all able to reach a cold
matter behavior before the present time. Hence, in the definition of
the matter spectrum we just write $\delta_m = (\delta \rho_b + \delta
\rho_\phi)/(\rho_b + \rho_\phi)$ without making any distinction about
the time before or after the onset of the field oscillations. One
consequence of this is that the matter spectrum is mainly determined
by the baryon perturbations at early times, but by the SFDM
perturbations at late times once the scalar field dominates the matter
sector.

\subsection{Modifications in CLASS}
\label{sec:modifications-class-1}
The numerical evolution of the scalar field perturbations within CLASS
is trickier than the evolution of the background, mainly because we
need to take care of the different time scales that explicitly appear in
the equations of motion, represented here by the angular variables
$\theta$, $\vartheta$, and their combinations $\theta \pm
\vartheta$. In order to make CLASS keep a good track of the evolution,
it is convenient to use the following equations of motion that result
from the combination of Eqs.~\eqref{eq:4a} and~\eqref{eq:25a}:
  \begin{subequations}
    \label{eq:33}
    \begin{eqnarray}
      \tilde{\vartheta}^\prime &=& -3 \left[ \sin \theta + \sin
        (\theta - \tilde{\vartheta}) \right] - \left[ 1 - \cos(
        \theta - \tilde{\vartheta}
        ) \right] \omega + e^{-\alpha} h^\prime \left[
        \cos(\tilde{\vartheta}/2)
        - \cos(\theta -
        \tilde{\vartheta}/2)
      \right] \, , \label{eq:33a}
      \\
      \alpha^\prime &=& -\frac{3}{2} \left[ \cos (\theta -
        \tilde{\vartheta}) + \cos \theta \right] - \frac{\omega}{2}
      \sin (\theta - \tilde{\vartheta})
      + \frac{1}{2} e^{-\alpha} h^\prime \left[
        \sin(\tilde{\vartheta}/2) +
        \sin(\theta - \tilde{\vartheta}/2) \right] \,
      , \label{eq:33c}
    \end{eqnarray}
\end{subequations}
where we have defined the new angular variable: $\tilde{\vartheta} = \theta - \vartheta$. As anticipated in
Sec.~\ref{sec:line-axion-pert}, variables $\theta$ and $\vartheta$
will always have a very similar evolution, and then variable
$\tilde{\vartheta}$, whose equation of motion~\eqref{eq:33a} does not
contain $y_1$, will just measure the small difference between the
two. Eqs.~\eqref{eq:33} together are the fundamental set of equations
of motion for scalar field perturbations that were implemented in
CLASS. The initial conditions for the new angular variables $(\tilde{\vartheta},\alpha)$ are inferred from the attractor conditions~\eqref{eq:45}, and then:
\begin{equation}
  \tilde{\vartheta}_i = (1/6) \theta_i \, , \quad \alpha_i =  \ln
  \left( h_i \theta_i /7\right) \, . \label{eq:24}
\end{equation}

In order to deal with the rapidly oscillating terms in the
perturbation equations, we also applied to them the cut-off procedure
of Eq.~\eqref{eq:18}, and then we made the replacements $\{ \cos,
\sin\} \to \{ \cos_\star, \sin_\star\}$ for all the trigonometric
functions in Eqs.~\eqref{eq:33} containing $\theta$ in the
argument. Like in the case of the background variables, those
trigonometric functions were numerically neglected once their
corresponding arguments reached a high-frequency regime.  For
  instance, the second term on the rhs of Eq.~\eqref{eq:33a} was
  written as $\sin_\star(\theta - \tilde{\vartheta})$, and then the
  cut-off to its evolution was applied once $\theta -
  \tilde{\vartheta} > 10^2$. Being $\theta$ the dominant angular variable,
  this also means that the rapidly oscillating terms are all cut-off
  at the same time for both the background and the perturbed equations
of motion.

\begin{figure*}[htp!]
\centering
\includegraphics[width=\textwidth]{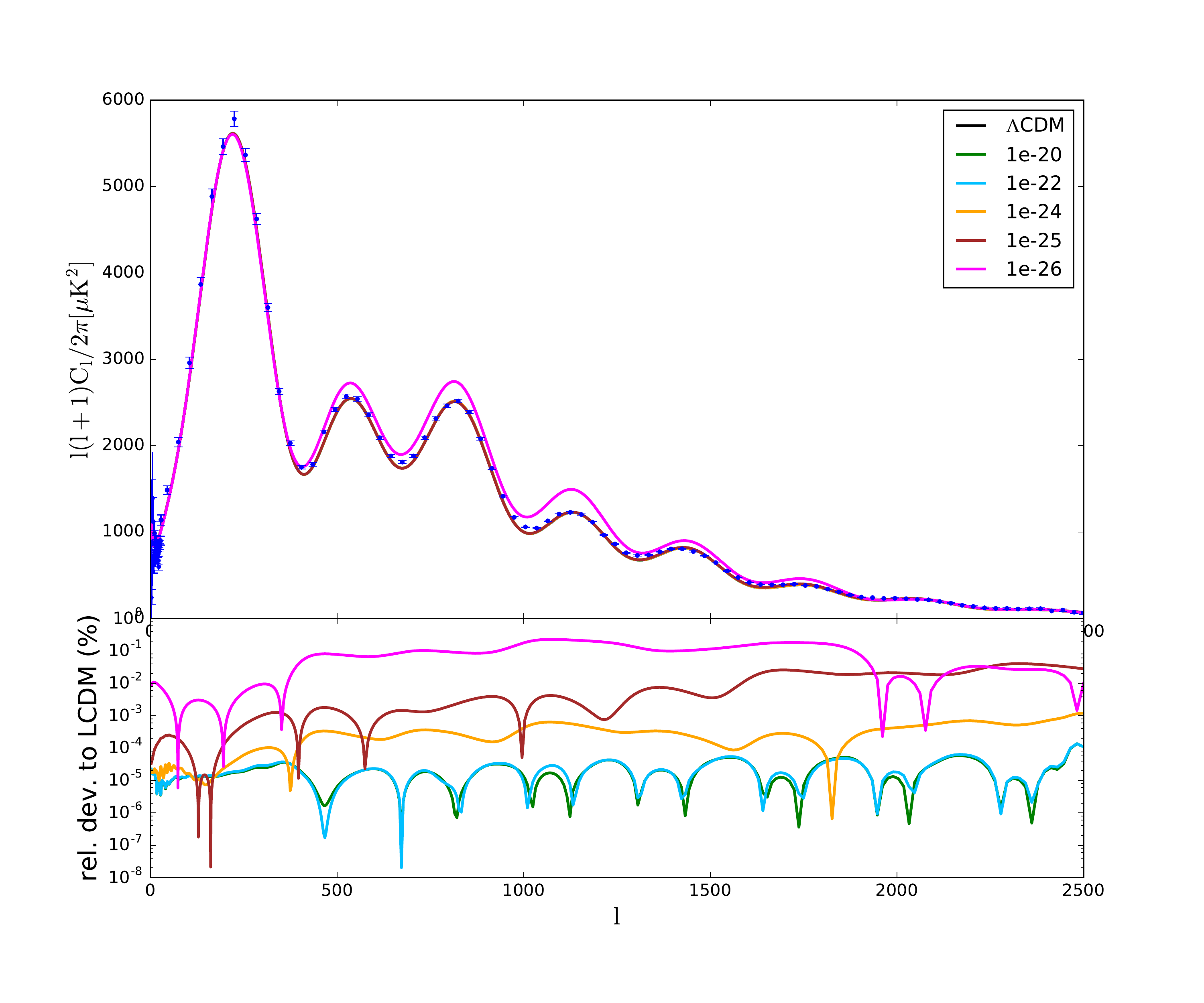}

\caption{\label{fig:3}CMB temperature spectra for $\Lambda$SFDM, as compared to the standard $\Lambda$CDM
  case. The differences are marginal for masses $m \gtrsim 10^{-24}
  {\rm eV}$, but they become quite noticeable for the lightest
  cases. Blue dots correspond to the Planck 2015 observations for binned high-l and low-l \cite{Ade:2015xua}}
\end{figure*}

\begin{figure*}[htp!]
\includegraphics[width=\textwidth]{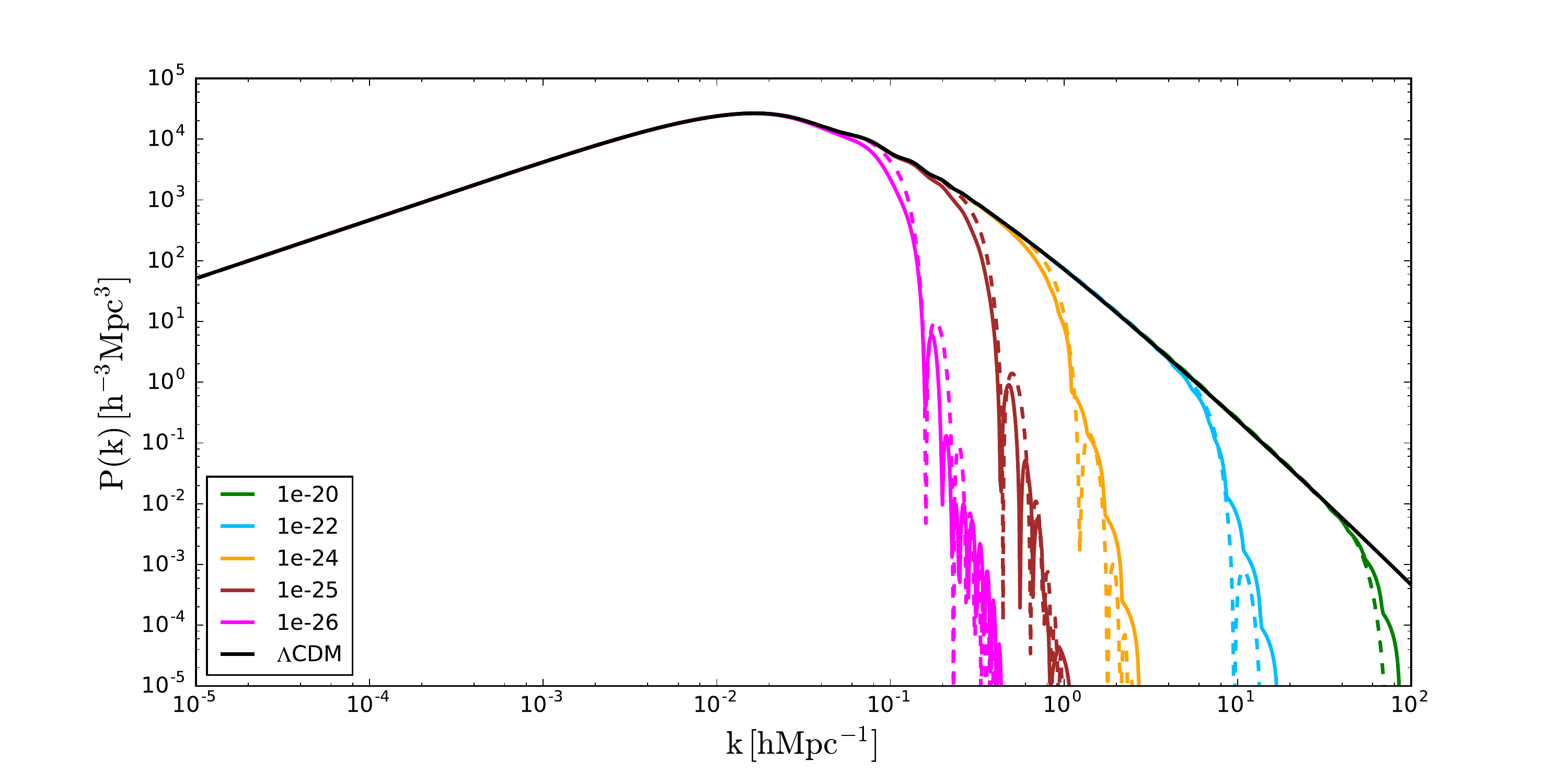}

\caption{\label{fig:PS_pk}  Mass PS for different
  masses of the scalar field, and for the $\Lambda$CDM model (black line), as obtained from the amended version of CLASS. 
The most noticeable effect is the well-known cut-off  at large scales (small k's) for the lightest mass considered.  Also shown for comparison is the approximated power spectrum using equation ~\eqref{eq:psShive} (dashed lines), for the corresponding scalar field mass. Qualitatively the two approaches are consistent, see the text for a more quantitative discussion.}
  
\end{figure*}

The equations of motion of perturbations were coded in CLASS following
the formalism of the synchronous gauge, which is permitted as long as
CDM is part of the matter budget. As mentioned before in
Sec.~\ref{sec:gauge-ambig-matt}, our simulations considered that
$\Omega_{CDM 0} = 10^{-6}$, so that CLASS could still be able to fix
the gauge completely. It must be remarked here that all SFDM
  physical quantities were written inside CLASS exactly as presented
  in the text, without any handling of them except for the cut-off of
  those trigonometric terms that were able to reach a high-frequency
  regime. For example, as part of the source terms in the perturbed
  Einstein equations, the SFDM density contrast was written exactly as
  $\delta_\phi = -e^\alpha \sin (\tilde{\vartheta}/2)$, whereas the
  pressure contrast was written as $\delta_{p_\phi} = - e^\alpha
  \sin_\star(\theta + \tilde{\vartheta}/2)$.

In Figs.~\ref{fig:3} and \ref{fig:PS_pk} we show the results from CLASS corresponding to
the CMB temperature and mass power spectra, for the same range of
masses as in the background dynamics: $10^{-26} < m/{\rm eV} <
10^{-20}$, and compare them with the standard $\Lambda$CDM model. In
general terms, SFDM perturbations resemble very much those of the
standard CDM plots, except for the lightest masses considered for
which differences are more than noticeable. As for the matter
spectrum, the most remarkable feature is the expected sharp cut-off
for the smallest scales, which is clearly seen in the cases $m
\lesssim 10^{-24} {\rm   eV}$.

One known approach to calculate the mass power spectrum for
the SFDM model was to rescale the $\Lambda$CDM one using the expression:
\begin{equation}
  P_{\rm SFDM}(k,z)=T^2(k) P_{\Lambda\rm{CDM}}(k,z)
\label{eq:psShive}
\end{equation}
where the transfer function $T(k)$ is effectively considered as redshift
independent:
\begin{equation}
T(k)=\frac{\cos (x^3)}{(1+x^8)} \, , \quad x= 1.61 \left(
  \frac{m}{10^{-22} {\rm eV}} \right)^{1/18} \frac{k}{k_{J,eq}} \, ,
\end{equation}
and the Jeans scale reads $k_{J,eq}=9 m_{22}^{1/2} \rm{Mpc}^{-1} $. In
Fig.~\ref{fig:PS_pk}, we show the SFDM PS calculated from
Eq.~\eqref{eq:psShive} and the one calculated numerically in this
work, and both of them are compared with the $\Lambda$CDM one. We can
notice that the cut-off seem to match well in both
approaches, and our result is in agreement with those already reported
in the literature
\cite{Hu:1998kj,Hu:2000ke,Matos:2000ng,Woo:2008nn,Marsh:2010wq,Hlozek:2014lca,Park:2012ru,Beyer:2014uja}. We
shall discuss the differences more in detail in Sec.~\ref{sec:mass-power-spectrum} below.

\section{Suppression of power in scalar field perturbations \label{sec:suppression-power-at}}
The common explanation for the suppression of power in the mass
spectrum of SFDM is usually given in terms of the Jeans wave
number~\eqref{eq:23}, but still this is a confusing subject in the
literature, mostly because of the different by-hand manipulations in
the fluid approximation of the equations of motion. To illustrate the
point, notice that we can write: $k^2/k^2_J = (k^2/a^2m^2)(m/H)$, and
then previous works have made separate and complicated studies of the
perturbations in terms of the ratios $k/am$ and
$m/H$\cite{Park:2012ru,Marsh:2010wq,Hlozek:2014lca,Alcubierre:2015ipa}. As
we are to explain now, the linear equations~\eqref{eq:33} provide a
simpler, though a bit subtler, explanation for the suppression of
linear perturbations. The semi-analytic results below will be compared
with the numerical ones obtained from the amended version of CLASS,
which are shown in Figs.~\ref{fig:8},~\ref{fig:6} and~\ref{fig:7}
below.

\subsection{General form of the equations of motion for scalar field
  perturbations \label{sec:gener-form-equat}}
For a better visualization of the physics behind the evolution of
scalar field perturbations, it is convenient to write the equations of
motion~\eqref{eq:33} in the following form:
\begin{eqnarray}
  \label{eq:41}
    \left[ 
\begin{array}{c}
e^\alpha \cos(\tilde{\vartheta}/2) \\
e^\alpha \sin(\tilde{\vartheta}/2)
\end{array}
\right]^\prime = -\left[ 
\begin{array}{cc}
  3\cos \theta + \tilde{\omega} \sin \theta & \tilde{\omega} (1
                                                 + \cos \theta) \\
  3\sin \theta + \tilde{\omega} (1 - \cos \theta) & - \tilde{\omega}
                                                    \sin \theta
\end{array} \right] %\mathbf{P}
\left[ 
\begin{array}{c}
e^\alpha \cos(\tilde{\vartheta}/2) \\
e^\alpha \sin(\tilde{\vartheta}/2)
\end{array}
\right] + \frac{h^\prime}{2} \left[ 
\begin{array}{c}
\sin \theta \\
1-\cos \theta
\end{array}
\right] \, .
\end{eqnarray}
where $\tilde{\omega} = \omega/2$. To find the solutions of the field
perturbations at early times, before the onset of the field
oscillations, we just set $\theta = 0$ in the $2\times 2$ matrix, and
retain the lowest order terms of $\theta$ in the last vector on the
rhs of Eq.~\eqref{eq:41}. The latter then becomes:
\begin{equation}
\label{eq:43}
\left[ 
  \begin{array}{c}
    e^\alpha \cos(\tilde{\vartheta}/2) \\
    e^\alpha \sin(\tilde{\vartheta}/2)
  \end{array}
\right]^\prime \simeq - \left[ 
  \begin{array}{cc} 
    3 & \tilde{\omega} \\
    0 & 0
  \end{array} \right] \left[ 
  \begin{array}{c}
    e^\alpha \cos(\tilde{\vartheta}/2) \\
    e^\alpha \sin(\tilde{\vartheta}/2)
  \end{array}
\right] + \frac{h^\prime}{2} \left[ 
  \begin{array}{c}
    \theta \\
    \theta^2/2
  \end{array}
\right] \, .
\end{equation}
If we again consider that during RD: $h = h_i (a/a_i)^2$ and $\theta =
\theta_i (a/a_i)^2$, the growing solutions of Eq.~\eqref{eq:43} are:
\begin{equation}
  \left[ 
  \begin{array}{c}
    e^\alpha \cos(\tilde{\vartheta}/2) \\
    e^\alpha \sin(\tilde{\vartheta}/2)
  \end{array}
\right] \simeq h \theta \left[ 
  \begin{array}{c}
    1/7 \\
    \theta/12
  \end{array}
\right] \, .
\end{equation}

In agreement with the results in Sec.~\ref{sec:early-time-behavior}, inhomogeneities in the scalar field are then negligible at early
times, mostly because $\theta$ diminishes the influence of the driving
term in Eq.~\eqref{eq:41}. This situation can only be changed once the
field approaches the onset of oscillations at $\theta = \pi/2$. Hence,
we see that one reason for the suppression of power in the matter
spectrum is the late appearance of the oscillations of the scalar
field, and clearly such suppression is more acute for smaller values of
the field mass $m$. This is a general effect that happens for all
scales, as Eq.~\eqref{eq:43} is independent on the wavenumber $k$.

\subsection{General solution of perturbations after the onset of the
  field oscillations \label{sec:gener-solut-pert}}
Once the field oscillations start, any trigonometric function that
depends on $\theta$ must be considered a \emph{rapidly oscillating}
function. Following our cut-off procedure\footnote{The influence
    of the rapidly oscillating terms in the solution of
    Eq.~\eqref{eq:41} is explained in the
    Appendix~\ref{sec:linear-perturbations}, where it is shown that
    the solution at late times coincides with that of
    Eq.~\eqref{eq:32}.} (see
  Eq.~\eqref{eq:18}), we shall consider that: $\sin_\star\theta \to 0$
  and $\cos_\star \theta \to 0$. Without such rapidly-oscillating
trigonometric functions, Eq.~\eqref{eq:41} now becomes:
\begin{subequations}
\label{eq:28}
\begin{equation}
  \label{eq:31}
  \left[ 
    \begin{array}{c}
      e^\alpha \cos(\tilde{\vartheta}/2) \\
      e^\alpha \sin(\tilde{\vartheta}/2)
    \end{array}
  \right]^\prime \simeq \tilde{\omega} \left[ 
\begin{array}{cc}
0 & 1 \\
-1 & 0
\end{array} \right] \left[ 
\begin{array}{c}
e^\alpha \cos(\tilde{\vartheta}/2) \\
e^\alpha \sin(\tilde{\vartheta}/2)
\end{array}
\right]
+ \frac{h^\prime}{2} \left[ 
\begin{array}{c}
0 \\
1
\end{array}
\right] \, .
\end{equation}
Eq.~\eqref{eq:31} has the form of a forced harmonic oscillator with
characteristic frequency $\tilde{\omega}$. In order to find
approximate solutions, we will assume that $\tilde{\omega} = {\rm
  const.}$ (which is only strictly true during radiation domination),
and then Eq.~\eqref{eq:31} formally integrates into:
\begin{equation}
  \left[ 
\begin{array}{c}
e^\alpha \cos(\tilde{\vartheta}/2) \\
e^\alpha \sin(\tilde{\vartheta}/2)
\end{array}
\right] \simeq \left[ \begin{array}{cc}
\cos(\tilde{\omega} N) & \sin(\tilde{\omega} N) \\
-\sin(\tilde{\omega} N) & \cos(\tilde{\omega} N)
\end{array} \right] \left[ \begin{array}{c}
C_1 \\
C_2
\end{array} \right] + \frac{1}{2} \int^N_0 d\hat{N} h^\prime \left[ \begin{array}{c}
\sin[\tilde{\omega} (N-\hat{N})] \\
\cos[\tilde{\omega} (N-\hat{N})]
\end{array} \right] \, , \label{eq:32}
\end{equation}
where $C_1,C_2$ are integration constants, and $N =\ln
(a/a_{osc})$. The integral on the rhs of Eq.~\eqref{eq:32} can be made
only if we know the form of the metric continuity $h^\prime$, which in
turn can only be found after solving the full set of perturbation
equations for all the matter fields.

As we are going to show now, a shortcut is to assume that we somehow
know the behavior of the metric term $h$. In general, we will have two
possibilities. The first one is to consider that $h = h_0 e^{\beta
  N}$, where $\beta >0$ is a constant, and $h_0$ is the initial value
of $h$. For instance, in the standard $\Lambda$CDM model we find that
the growing solutions of the metric perturbations are of the form $h =
h_0 e^{4N}$ ($h= h_0 e^N$) during RD (MD), and then we can assume that
$h^\prime = \beta h_0 e^{\beta N}$. Likewise, in the case that matter perturbations are suppressed\footnote{Such behavior of the
metric perturbation $h$ is expected if the perturbed Einstein
equations are not sourced at all by any density perturbations, see for
instance Eqs.~(21) in\cite{Ma:1995ey}.} we then expect that $h \simeq
h_\infty + (h_0 - h_\infty) e^{\beta N}$, here $h_0$ ($h_\infty$) are
the values at $N \to 0$ ($N \to \infty$); in this case $\beta < 0$ and
then $h^\prime = \beta (h_0 - h_\infty) e^{\beta N}$. The general
solution of Eq.~\eqref{eq:32}, in either of the aforementioned cases,
can be written as:
\begin{equation}
  \left[ 
    \begin{array}{c}
      e^\alpha \cos(\tilde{\vartheta}/2) \\
      e^\alpha \sin(\tilde{\vartheta}/2)
    \end{array}
  \right] \simeq 
  \left[ \begin{array}{cc}
           \cos(\tilde{\omega} N) & \sin(\tilde{\omega} N) \\
           -\sin(\tilde{\omega} N) & \cos(\tilde{\omega} N)
         \end{array} \right] 
       \left[ \begin{array}{c}
                C_1 - \frac{\tilde{\omega}}{2(\beta^2 +
                \tilde{\omega}^2)} h^\prime_0 \\
                C_2 - \frac{\beta}{2(\beta^2 + \tilde{\omega}^2)} h^\prime_0 
              \end{array} \right] 
            + \frac{h^\prime}{2(\beta^2 + \tilde{\omega}^2)}
            \left[ \begin{array}{c}
                     \tilde{\omega} \\
                     \beta
                   \end{array} \right] \, . \label{eq:20}
\end{equation}
\end{subequations}

\begin{figure*}[htp!]
\includegraphics[width=0.49\textwidth]{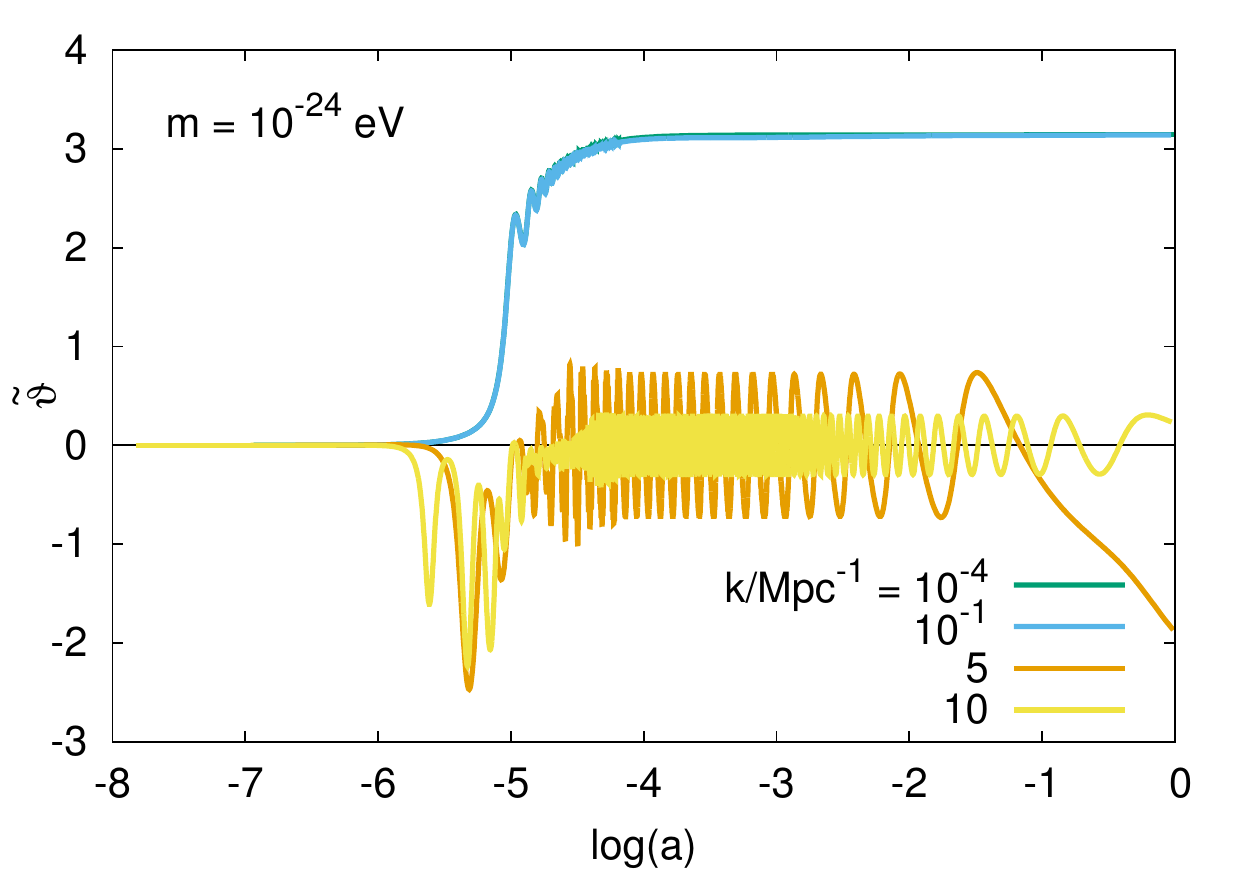}
\includegraphics[width=0.49\textwidth]{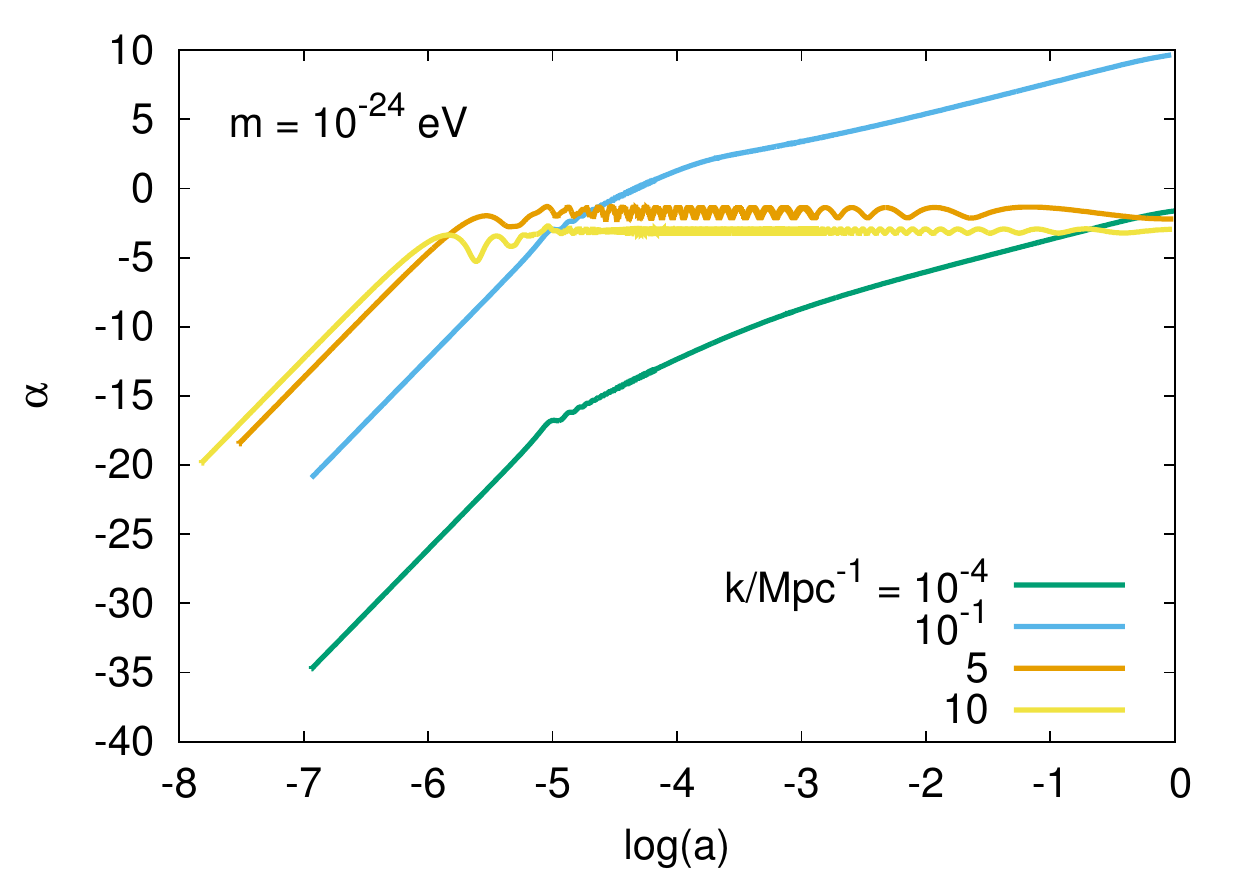}
\caption{\label{fig:8}Numerical solutions for the perturbation
  variables $\tilde{\vartheta}$ (left) and $\alpha$ (right) for an
  example with a mass $m=10^{-24} \, {\rm eV}$. For each variable we
  show two \emph{growing} and two \emph{decaying} cases, following the
  definitions in Sec.~\ref{sec:gener-solut-pert}, which
  correspond to the wavenumbers $k= 10^{-4}, 10^{-1} \, {\rm
    Mpc}^{-1}$, and $k= 5, 10 \, {\rm Mpc}^{-1}$, respectively. The Jeans wavenumber for the given mass is $ k_{J0} =1.8 {\rm Mpc}^{-1}$, see Eq.~\eqref{eq:23b}, and
  then the cases can also be distinguished by the value of the
  frequency $\omega$: for the growing case $\omega \ll 1$, whereas
  $\omega \gg 1$ for the decaying one. In general, we can observe that
  for the growing case $\tilde{\vartheta} \to \pi$ (note that both green and
  blue lines coincide on the left panel) and $\alpha$ grows, whereas
  for the decaying case both $\tilde{\vartheta}$ and $\alpha$ become
  oscillatory functions. These behaviors are in very good
    agreement with those inferred from the semi-analytical results in
    Eqs.~\eqref{eq:28} and~\eqref{eq:40}. In particular, we can see by
    eye that the solutions for small scales oscillate in terms of the
    number of $e$-folds $N$. However, the numerical solutions show
    that the frequency of the oscillations decreases as the evolution
    proceeds, which is expected as $\tilde{\omega}$ decreases after
    the epoch of RD. See the text for more details.}
\end{figure*}

The general solution then consists of an oscillating part plus an
extra term proportional to the driving term containing the metric
perturbation. Before studying the general cases, we analyze the forced
solutions without the oscillating terms. The separate solutions of the
variables $\tilde{\vartheta}$ and $e^\alpha$ are, in full generality,
given by:
\begin{subequations}
\label{eq:40}
\begin{equation}
  \tan(\tilde{\vartheta}/2) \simeq \frac{\beta}{\tilde{\omega}} \, ,
  \quad e^\alpha \simeq \frac{h^\prime}{2(\beta^2 +
    \tilde{\omega}^2)^{1/2}} \, . \label{eq:40a}
\end{equation} 
As we shall see now, there are only two \emph{consistent} solutions if
the scalar field perturbations are going to be the main source of
matter perturbations.

The first one is what we call the \emph{growing} case, which is
achieved if $\omega \ll 1$ and $e^\alpha$ is a growing function. From
the discussion above, this case corresponds to $h^\prime = \beta h$,
and then:
\begin{equation}
\label{eq:34}
    \tilde{\vartheta} \simeq \pi - 2\tilde{\omega}/\beta \, , \quad
    e^\alpha \simeq \frac{1}{2} h (1-\tilde{\omega}^2/\beta^2) \, ,
\end{equation}
where we have included the correction at the lowest order in
$\tilde{\omega}/\beta \ll 1$. The first thing to notice is that the angular
variable $\tilde{\vartheta} \to \pi$, and also $e^\alpha \to h/2$. In consequence, the scalar field density contrast at late times,
from Eq.~\eqref{eq:26a}, can be written as $\delta_\phi \simeq -(h/2)
\cos(\tilde{\omega}/\beta) \simeq -h/2$  (notice that this
  resembles the functional form of the CDM growing mode solution,
  which is $\delta_{\rm CDM} = -h/2$ exactly). The solution is
consistent in the sense that both the field perturbations and the
metric continuity are growing functions. The numerical solutions
corresponding to this case are shown in Fig.~\ref{fig:8}, for the
wavenumbers $k=10^{-4}, 10^{-1} \, {\rm Mpc}^{-1}$ and a scalar field
mass $m=10^{-24} \, {\rm eV}$.

The second possibility is the \emph{decaying} case, for which
$\tilde{\omega} \gg 1$ and $e^\alpha$ is a non-growing function. The
latter condition can be achieved in the second case discussed above
corresponding to $h^\prime = \beta (h_0 - h_\infty)e^{\beta N}$, from
which we obtain that:
\begin{equation}
\label{eq:12}
 \tilde{\vartheta} \simeq 2 \beta/\tilde{\omega} \, , \quad e^\alpha
 \simeq \frac{\beta/\tilde{\omega}}{2(1 +
   \beta^2/\tilde{\omega}^2)^{1/2}} (h_\infty -h) \, ,
\end{equation}
\end{subequations}
where we included again the correction at the lowest order in
$\beta/\tilde{\omega} \ll 1$. Eq.~\eqref{eq:12} suggests that this
time the angular variable $\tilde{\vartheta} \simeq 0$ for
$|\beta/\tilde{\omega}| \ll 1$, and that the amplitude of
perturbations vanishes, $e^\alpha \to 0$, as $h \to h_\infty$. The two
results together imply that $\delta_\phi \to 0$. Like in the growing
case, the new solution is also consistent in the sense that both the
metric and field perturbations are non-growing functions.

However, as the driving term in Eq.~\eqref{eq:20} vanishes at late
times for the decaying case, the true solution is eventually
represented by its oscillatory part only. This is shown by the
numerical solutions in Fig.~\ref{fig:8} for the wavenumbers $k=5, 10
\, {\rm Mpc}^{-1}$, as it is for these values of $k$ that $\omega \gg
1$ if $m=10^{-24} \, {\rm eV}$. As a final note, we must mention
  that the case with $k=5 \, {\rm Mpc}^{-1}$ shows that
  $\tilde{\vartheta}$ slowly drifts towards $-\pi$ at late
  times\footnote{Notice that $\tilde{\vartheta} \to - \pi$, instead of
    $\tilde{\vartheta} \to \pi$, in the case $k= 5 \, {\rm
      Mpc}^{-1}$. This is not but a numerical glitch because CLASS
    cannot in this case keep an accurate calculation of the
    oscillations of $\tilde{\vartheta}$ at late times.}. This
is to be expected, given that our previous assumption that $\omega =
{\rm const.}$ is not strictly attained during the whole
evolution. Actually, we can see that the frequency of the oscillations
(with respect to the variable $N$) diminishes at late times, just
because the Jeans wavenumber $k_J$ is a growing function after the time of
radiation-matter equality (see the discussion around
Eq.~\eqref{eq:23a}). This is an interesting example in which one given
scale starts in the regime $\tilde{\omega} \gg 1$, but changes to the
one with $\tilde{\omega} \ll 1$ at late times.

Apart from the dynamical variables $\alpha,\tilde{\vartheta}$, there
are two quantities that are of more physical interest: the scalar
field density contrast $\delta_\phi$, and the velocity gradient
$\Theta_\phi$.  According to Eq.~\eqref{eq:26a}, the density
  contrast can be written as $\delta_\phi = - e^\alpha
  \sin(\tilde{\vartheta}/2)$, and then its general evolution can be
  obtained directly from the solutions presented above. To determine
  the general behavior of the velocity gradient $\Theta_\phi$, let us write
  Eq.~\eqref{eq:26c} in the form
  \begin{equation}
    \label{eq:65}
    (\rho_\phi + p_\phi) \Theta_\phi = \rho_\phi \Theta_\phi (1
    -\cos\theta) = -\frac{k^2}{am} \rho_\phi e^\alpha \sin(\theta/2)
    \sin(\vartheta/2) \, ,
  \end{equation}
where we have used the EoS from Eq.~\eqref{eq:6}. We can see that the
source of SFDM velocity perturbations is indeed a rapidly oscillating
function at late times. As we shall see, though, the important point
is to determine whether those oscillations are relevant or not by
calculating their amplitude, which is represented by the product
$\rho_\phi \Theta_\phi$. At early times we can write Eq.~\eqref{eq:65}
as:
\begin{subequations}
\begin{equation}
  \label{eq:66}
  \frac{1}{2} \rho_\phi \Theta_\phi \theta^2 \simeq -\frac{5 k^2}{24
    am} \rho_\phi e^\alpha \theta^2 \quad \to \quad \rho_\phi
  \Theta_\phi \simeq -\frac{5 k^2}{12 am} \rho_\phi e^\alpha \, ,
\end{equation}
whereas at late times, after the onset of the field oscillations, we
can use that $\vartheta \simeq \theta = 2mt$ and then
Eq.~\eqref{eq:65} can be written as:
\begin{equation}
  \label{eq:68}
  2 \rho_\phi \sin^2(mt) \Theta_\phi \simeq -\frac{k^2}{am} \rho_\phi
  e^\alpha \sin^2(mt) \quad \to \quad  \rho_\phi \Theta_\phi \simeq
  -\frac{k^2}{2 am} \rho_\phi e^\alpha \, .
\end{equation}
\end{subequations}
Hence, we will take the last term in Eq.~\eqref{eq:68} as a good
approximation at all times for the product $\rho_\phi \Theta_\phi$.
In the following sections, we will analyze the behavior of the
perturbation quantities $\delta_\phi$ and $\rho_\phi \Theta_\phi$ for
large and small scales as compared to the Jeans wavenumber $k_J$. 

\begin{figure*}[htp!]
\centering
\includegraphics[width=0.45\textwidth]{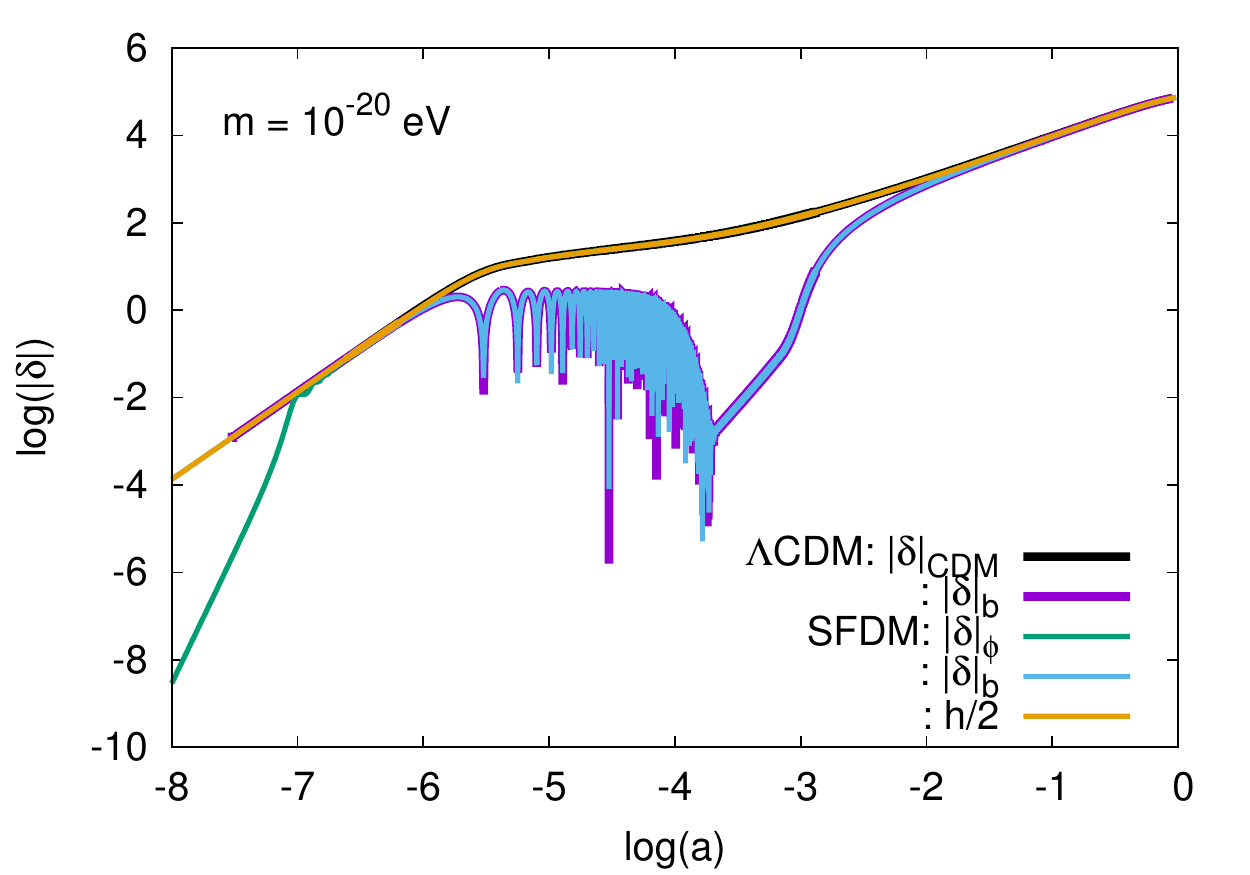}
\includegraphics[width=0.45\textwidth]{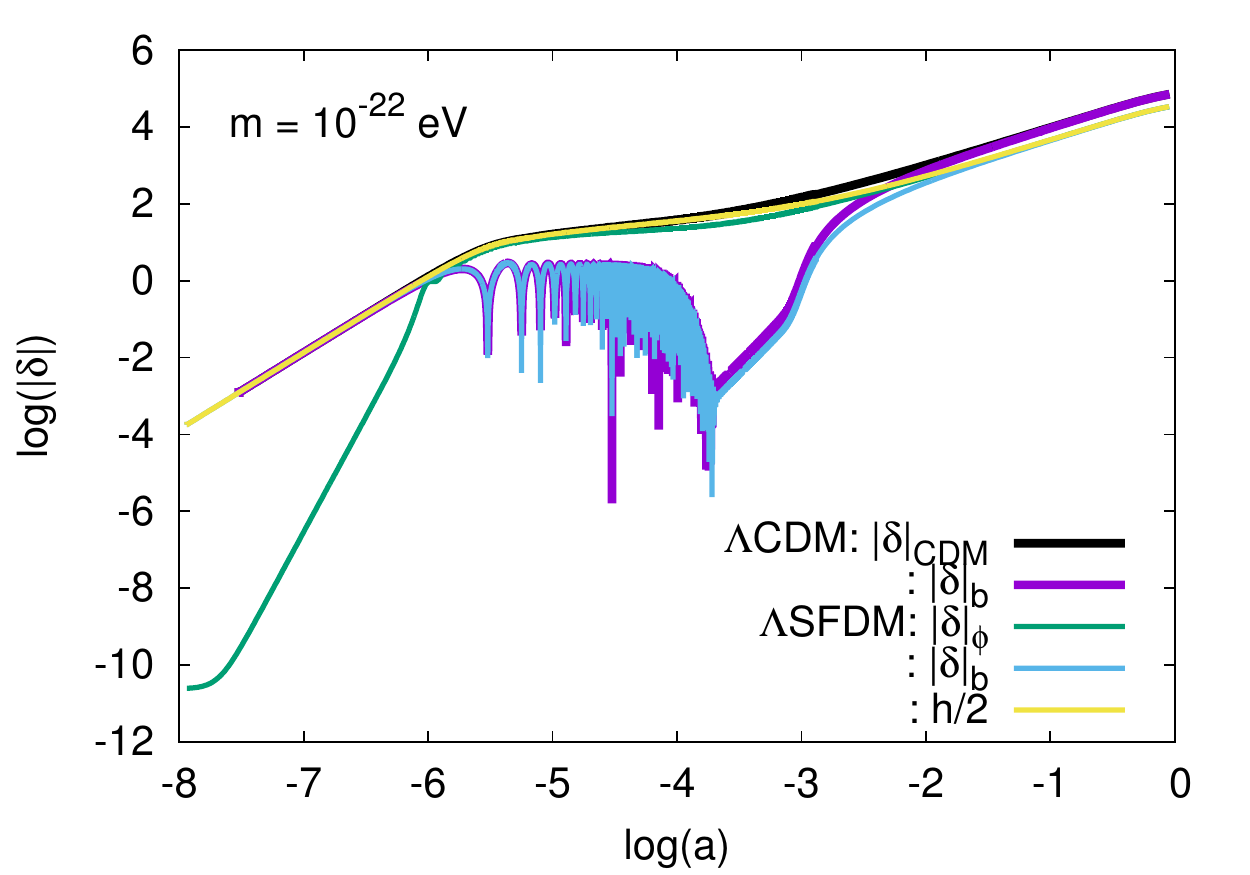}
\includegraphics[width=0.45\textwidth]{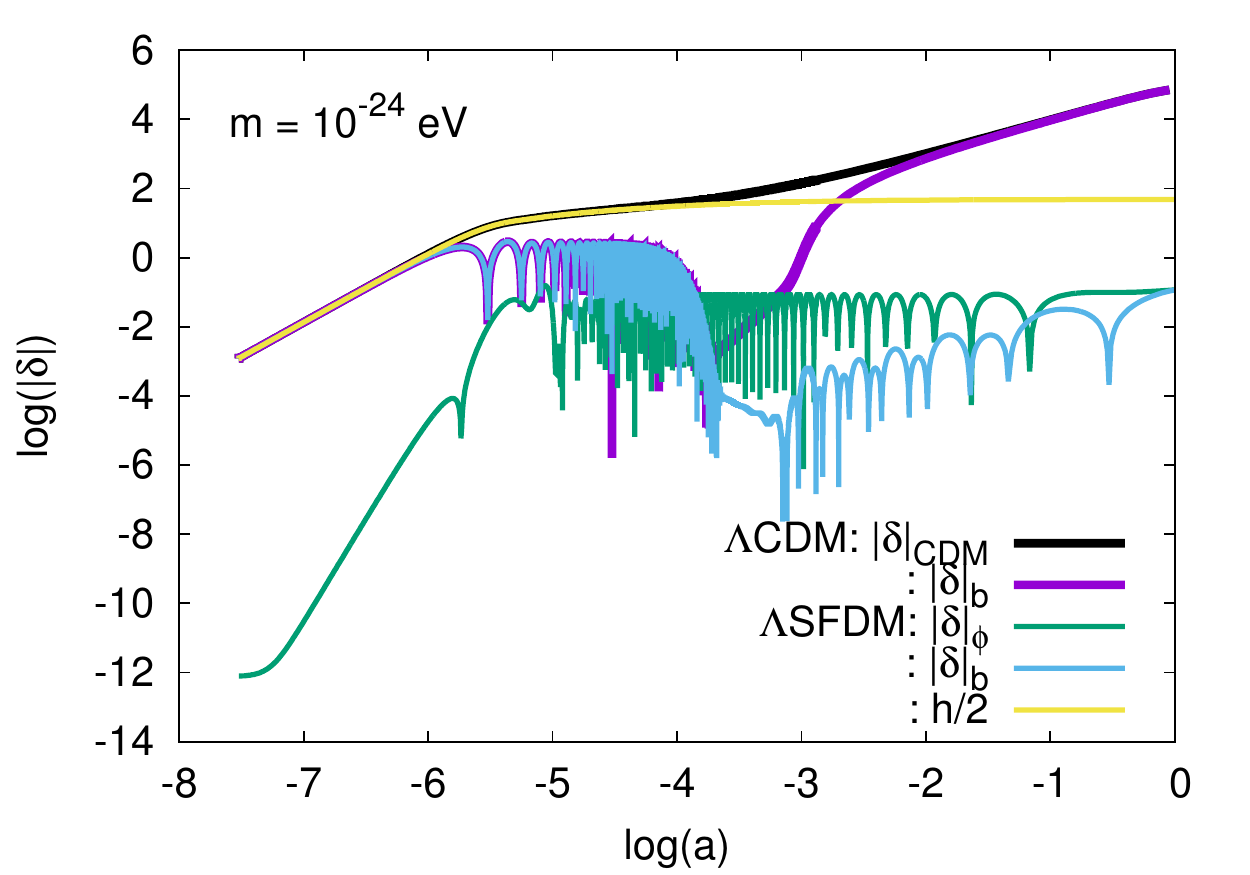}
\includegraphics[width=0.45\textwidth]{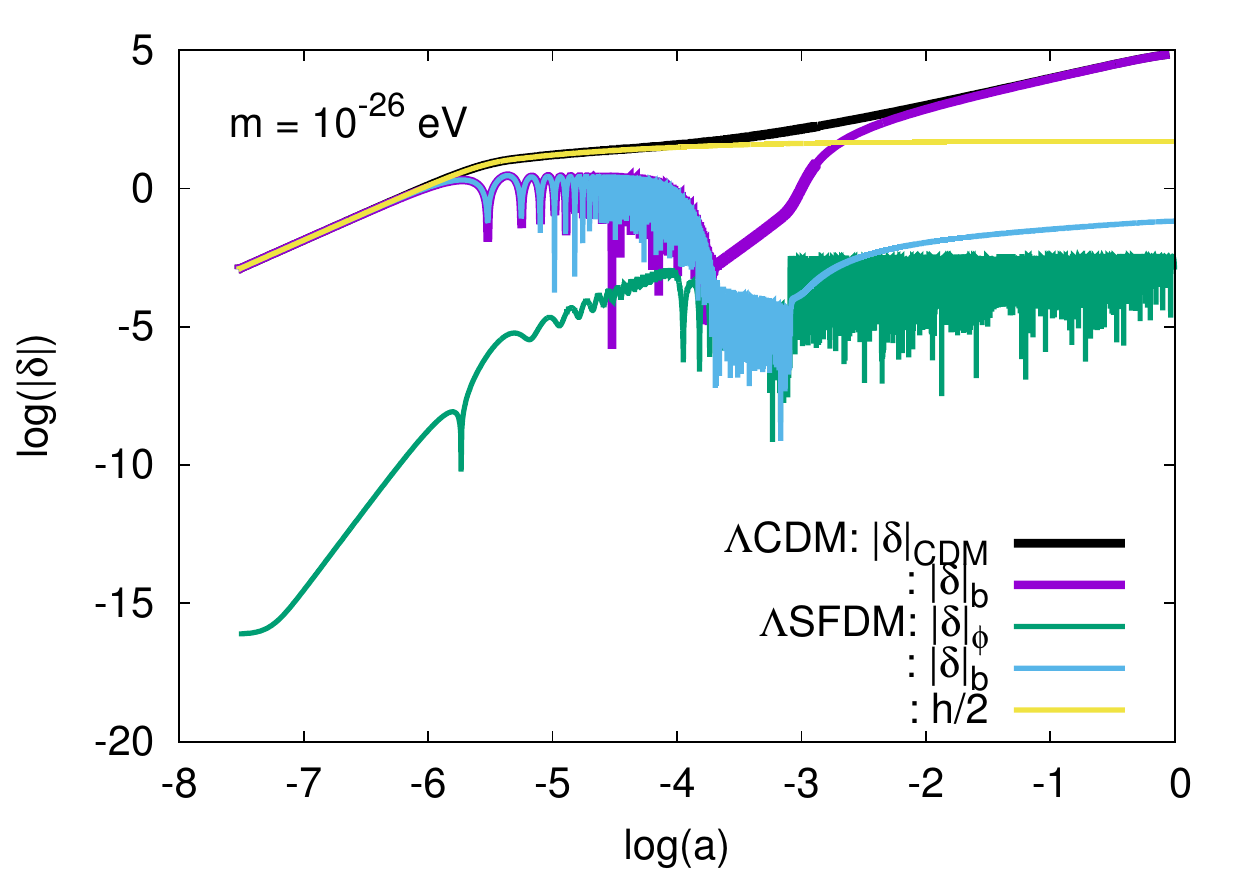}
\caption{\label{fig:6}The evolution of the density contrast of CDM,
  baryons ($b$), and SFDM ($\phi$) for the same wavenumber $k = 5 {\rm
    Mpc}^{-1}$, as obtained from the amended version of CLASS for the
  different scalar field masses indicated in each one of the
  plots. As explained in the Secs.~\ref{sec:large-scales-k}
  and~\ref{sec:small-scales-k} below, the important parameter for the
  suppression of the scalar field perturbations is the ratio $\tilde{\omega} =
  k^2/2k^2_J$, and then we refer to \emph{large} (\emph{small}) scales
  to the cases in which $\tilde{\omega} \ll 1$ ($\tilde{\omega} \gg
  1$). Given that this ratio depends on the mass of the scalar field,
  one same scale can give $\tilde{\omega} \ll 1$ for large masses (e.g. $m =
  10^{-20}, 10^{-22} {\rm eV}$), or $\tilde{\omega} \gg 1$ for small masses
  (e.g. $m = 10^{-24}, 10^{-26} {\rm eV}$). The latter case also
    means that there are some residual oscillations with frequency
    $\tilde{\omega}$ apart from the high-frequency ones related to
    $\theta$ that were cut-off in the modified version of CLASS, see
    Eq.~\eqref{eq:44} and Fig.~\ref{fig:8}}. In general, SFDM
  perturbations evolve as those of CDM for the large scales (see
  Sec.~\ref{sec:large-scales-k}), whereas they appear suppressed for
  the small scales (see Sec.~\ref{sec:small-scales-k}). Also shown is
  the evolution of the metric perturbation $h/2$ (see
  Eq.~\eqref{eq:13}), which for late times accomplishes the relation
  $\delta_\phi = -h/2$ for large scales, whereas for small scales it
  approaches an asymptotic value: $h \to h_\infty$. This is to be
  compared with the usual result $\delta_{\rm CDM} = - h/2$, which is
  exact for all scales in the case of CDM\cite{Ma:1995ey}. See the
  text for more details.
\end{figure*}

\begin{figure*}[htp!]
\centering
\includegraphics[width=0.45\textwidth]{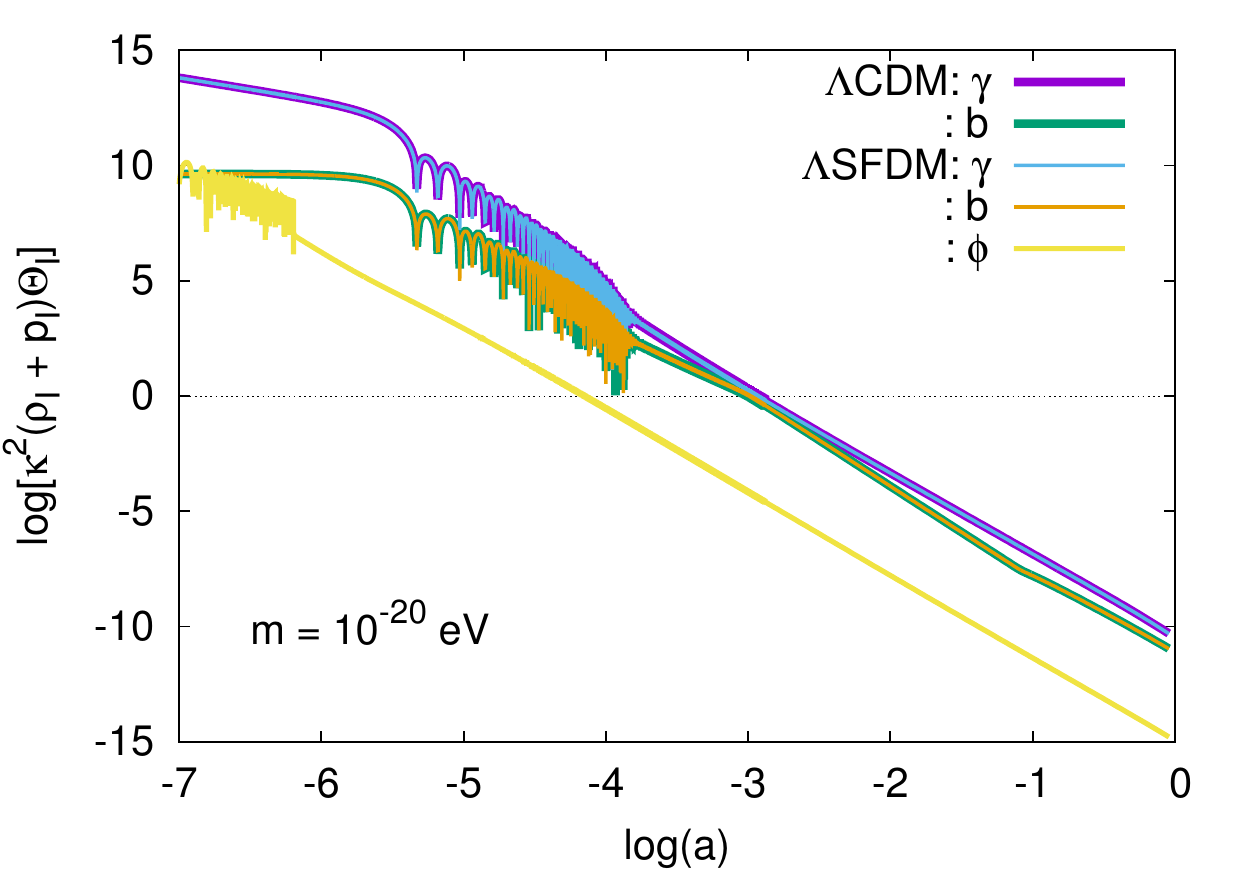}
\includegraphics[width=0.45\textwidth]{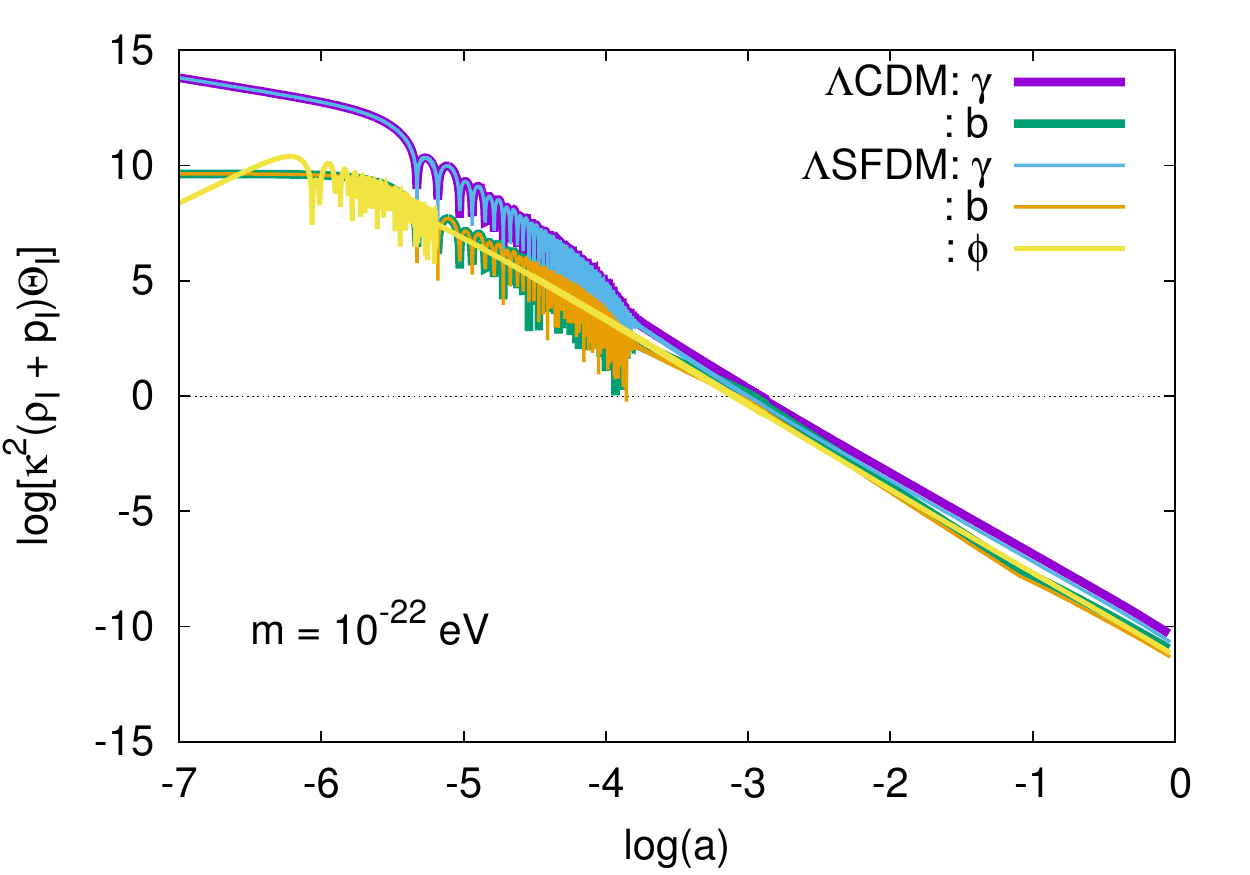}
\includegraphics[width=0.45\textwidth]{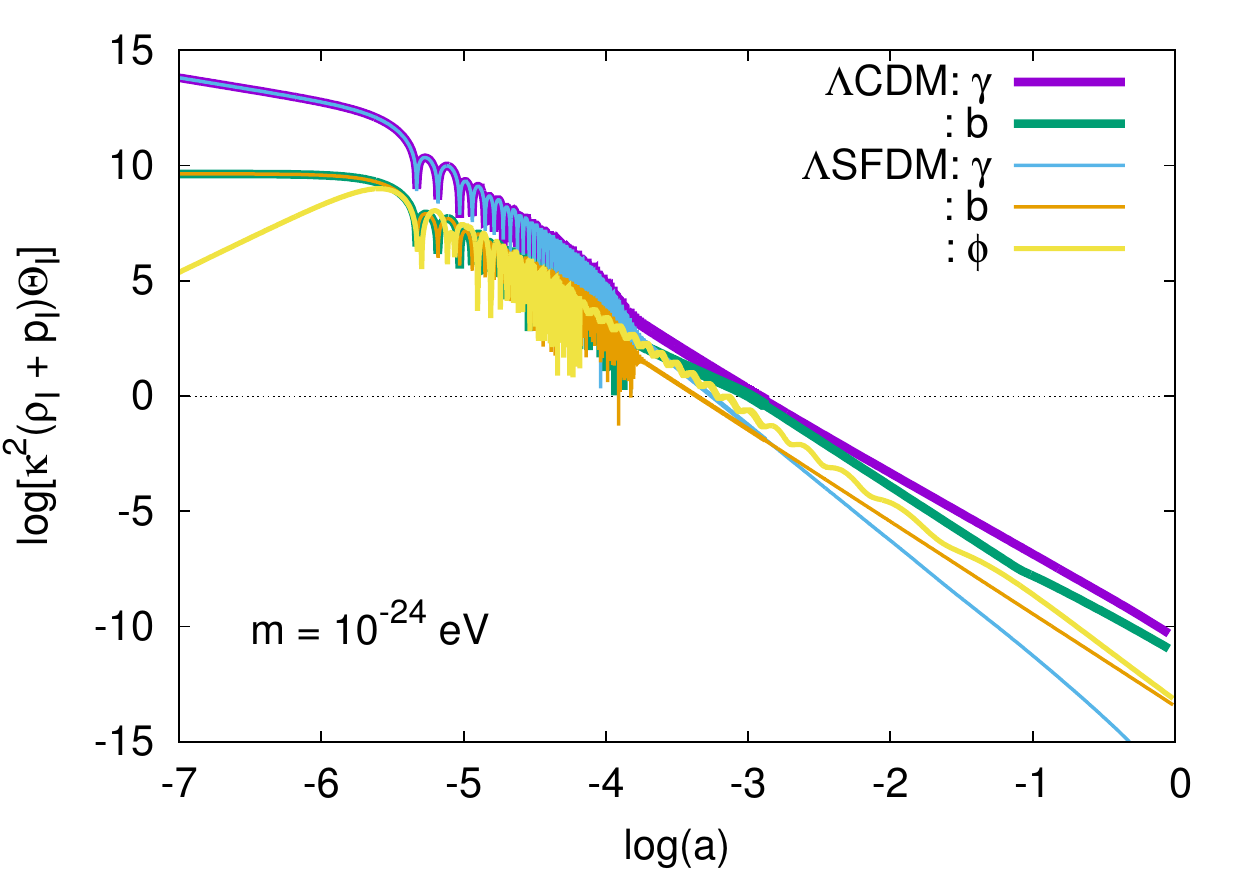}
\includegraphics[width=0.45\textwidth]{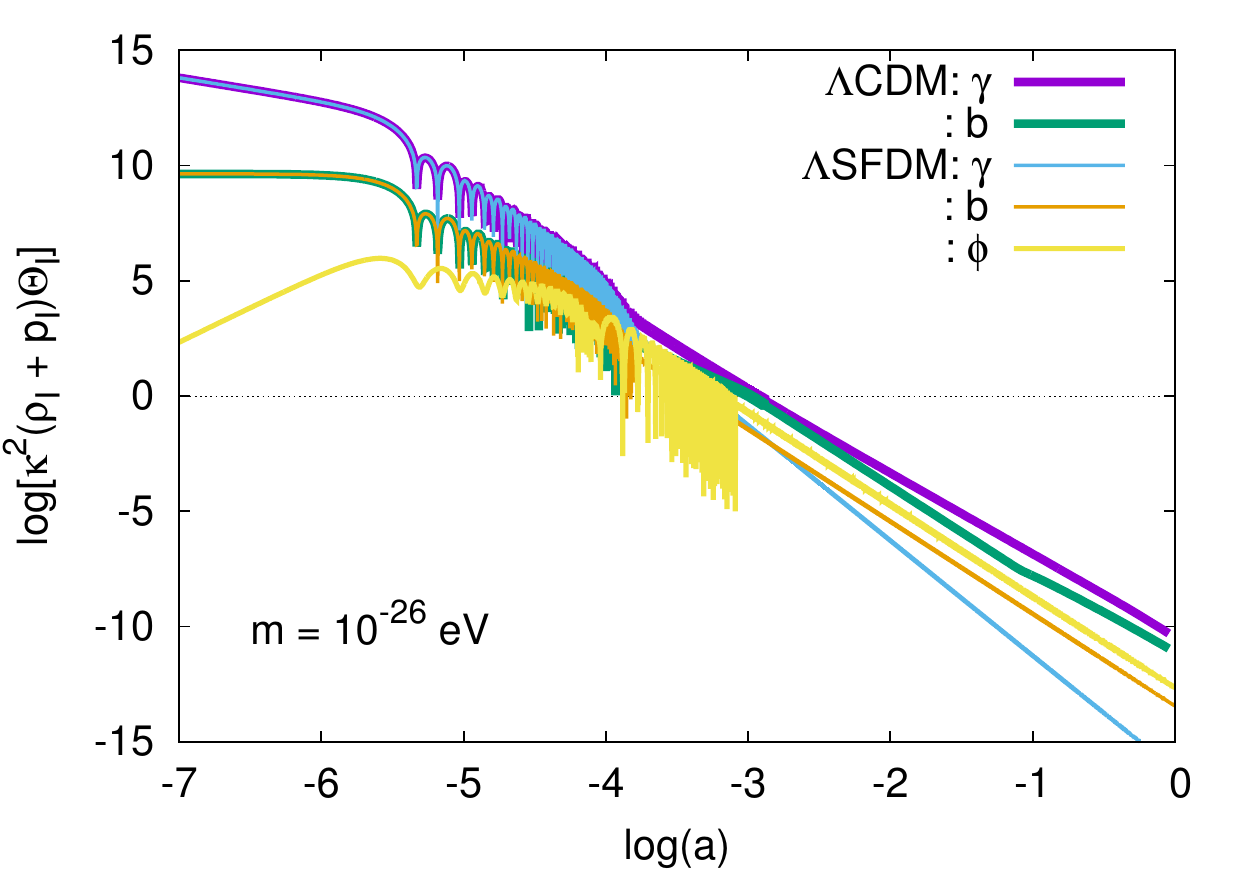}

\caption{\label{fig:7}The amplitude of the velocity perturbation
  source term $(\rho_I + p_I) \Theta_I$ of different matter components
  for the same wavenumber $k= 5 \, {\rm Mpc^{-1}}$ as in
  Fig.~\ref{fig:6}. That of a CDM component is not shown, as it is
  zero by definition in the synchronous gauge. The velocity gradient
  of photons and baryons, in the presence of SFDM, is the same as in
  $\Lambda$CDM for large scales  (top figures, for $m=
    10^{-20},10^{-22} {\rm eV}$), but they seem to decay more rapidly
    for small scales (bottom figures, for $m= 10^{-24},10^{-26} {\rm
      eV}$) at late times. It can be confirmed that SFDM velocity
    perturbations evolve as expected of a pressureless component in
    the frame comoving with a CDM component, which is indeed the gauge
    considered within CLASS for the calculation of perturbations. As
  a side effect, notice that velocity gradient of the photons is
  further suppressed at late times, as compared to that of the
  $\Lambda$CDM model, which is an expected result if $h^\prime \to 0$
  (see Eqs.~(29) in\cite{Ma:1995ey}). For more details, see the text
  in Secs.~\ref{sec:gauge-ambig-matt}, \ref{sec:large-scales-k}, and~\ref{sec:small-scales-k}.}
\end{figure*}

\subsection{Large scales $k \ll k_J$ \label{sec:large-scales-k}}
For large enough scales $k \ll k_J$, the characteristic frequency of
the oscillator~\eqref{eq:31} is small at all times: $\omega \ll
1$. According to our discussion above, the only \emph{consistent
  solution} for this case is the growing one represented by
Eq.~\eqref{eq:34}, and then the complete solution for the scalar field
density contrast, from Eq.~\eqref{eq:20}, is:
\begin{subequations}
\label{eq:35}
\begin{equation}
  \delta_\phi \simeq - \left[ C_1 - \frac{\beta^2}{2(\beta^2 +
  \tilde{\omega}^2)} h_0 \right] \sin(\tilde{\omega} N) - \left[ C_2 -
  \frac{\beta \tilde{\omega}}{2(\beta^2 + \tilde{\omega}^2)} h_0
  \right] \cos(\tilde{\omega} N) - \frac{\beta^2}{2(\beta^2 + \tilde{\omega}^2)} h \, . \label{eq:35a}
\end{equation}
If in addition $\omega \ll \beta$, this solution indicates that at
late times, $\delta_\phi \simeq - h/2$, with some superimposed
oscillations of small amplitude. Thus, SFDM behaves effectively as a
pressureless component at the level of linear perturbations for large
scales. This can also be corroborated from the calculation of
the pressure contrast in Eq.~\eqref{eq:26b}: $\delta_{p_\phi} =
-e^\alpha \sin(\theta + \tilde{\vartheta}/2)$, from which we see that
it is a rapidly oscillating function that becomes zero after the
cut-off of the oscillations.

On the other hand, we see that the asymptotic solution of the
amplitude of SFDM velocity perturbations at late times is:
\begin{equation}
  \label{eq:35d}
  \rho_\phi \Theta_\phi \simeq - \frac{k^2}{4am}
  \frac{\tilde{\omega}}{\beta} h \rho_\phi \simeq - \frac{k^2h_{osc}}{4m
    a_{osc}} \frac{\tilde{\omega}}{\beta} \rho_\phi \simeq {\rm
    const.} \times \tilde{\omega} a^{-3} \, ,
\end{equation}
\end{subequations}
where we have used Eq.~\eqref{eq:34}, and the last term appears if
$h=h_{osc} e^N = h_{osc} (a/a_{osc})$, which is the expected
expression during matter domination.  Eq.~\eqref{eq:35d} then
  indicates that there must be a residual amplitude of the velocity
  gradient at late times, which is quite small because of the
  assumption here that  $\tilde{\omega} \ll \beta$ ($k \ll
  k_J$). Moreover, the residual amplitude decays as $\tilde{\omega}
  a^{-3}$, where we must take into account that $\tilde{\omega}$ also
  contributes with an extra factor of $a^{-1}H^{-1/2} \sim a^{-1/4}$
during matter domination.

It must be recalled that the condition for large scales $k \ll k_J$
depends upon the value of the scalar field mass. In Figs.~\ref{fig:6}
and~\ref{fig:7} we give numerical examples obtained from CLASS of the
behavior of the density contrast $\delta_\phi$ and of the velocity
perturbation source term $(\rho_\phi + p_\phi) \Theta_\phi$,
respectively, in comparison with those of other matter components. The
large scale case corresponds, in both figures, to the masses $m =
10^{-20}, 10^{-22} \, {\rm eV}$, and it can be clearly seen that both
physical quantities evolve as suggested by the approximate solutions~\eqref{eq:35}.

 We have said before that the semi-analytical results suggest that
  for large scales $\delta_\phi \simeq -h/2$, just as for any cold
  matter component. This does not mean, however, that the amplitude of
  the SFDM density contrast is the same as in the CDM case. This can
  be noticed in Fig.~\ref{fig:6} for $m=10^{-22} {\rm eV}$, in which
  case the given wavenumber is mildly larger than the Jeans length: $k
  > k_J$, and in consequence $\delta_\phi$ has a bit smaller amplitude
  than $\delta_{CDM}$.

  As for the velocity gradient shown in Fig.~\ref{fig:7}, the same
  case with $m=10^{-22} {\rm eV}$ shows a brief stage of rapid
  oscillation, which is also seen to disappear because of the cut-off
  imposed upon the rapidly oscillating trigonometric functions in our
  amended version of CLASS.

\subsection{Small scales  $k \gg k_J$\label{sec:small-scales-k}}
For the case of small enough scales, $k \gg k_J$, corresponding to
$\omega \gg 1$, the consistent solution at late times for the density
contrast, from Eq.~\eqref{eq:20}, is:
\begin{equation}
  \delta_\phi \simeq - \left[ C_1 - \frac{\tilde{\omega} \beta (h_0
                       -h_\infty) }{2(\beta^2 +
      \tilde{\omega}^2)} \right] \sin(\tilde{\omega} N) + \left[ C_2 -
    \frac{\beta^2  (h_0 -h_\infty)}{2(\beta^2 + \tilde{\omega}^2)} \right]
  \cos(\tilde{\omega} N) + \frac{\beta^2  (h_\infty -
  h)}{2(\beta^2 + \tilde{\omega}^2)} \, , \label{eq:38}
\end{equation}
In contrast to the case of large scales, this time the driving term
decreases as $h \to h_\infty$ and does not become the dominant part of
the solution. Instead, in the limit $\beta/\tilde{\omega} \ll 1$, we
find that:
\begin{equation}
  \delta_\phi \simeq - C_1 \sin(\tilde{\omega} N) + C_2
  \cos(\tilde{\omega} N) \, . \label{eq:44}
\end{equation}
Hence, the density contrast turns into a (rapidly) oscillating function of
$N$ with a small amplitude, with the latter inherited from the
transition to the rapidly oscillating phase of the scalar field. As
for the SFDM velocity perturbations, see Eq.~\eqref{eq:65}, because
$e^\alpha \to {\rm const.}$ their amplitude quickly decreases as $\rho_\phi
\Theta_\phi \simeq (k^2/2am) \rho_\phi e^\alpha \sim a^{-4}$.  Together
with the results in the previous section, this reinforces our argument
in Sec.~\ref{sec:gauge-ambig-matt} that the velocity gradient of SFDM
 evolves as expected of a pressureless component in the gauge
 corresponding to the comoving frame of a CDM component.

Numerical examples for small scales are also shown in
Figs.~\ref{fig:6} and~\ref{fig:7} for the masses $m = 10^{-24},
10^{-26} \, {\rm eV}$, and they agree very well  with the behavior
  inferred from the approximate solutions described just above: the
density contrast is just an oscillating function with a tiny
amplitude, whereas the velocity gradient oscillates with a rapidly
decaying amplitude.

\section{Power spectrum and mass function}
\label{sec:mass-power-spectrum}

In order to exploit the constraining power of future galaxy
surveys such as DESI\cite{Levi:2013gra} and LSST\cite{Ivezic:2008fe},
it will be necessary to perform an accurate modeling of the linear, and non
linear, matter PS of any given dark matter model, for wavenumbers up to $k
\sim 10 h \, {\rm Mpc}^{-1}$. For $\Lambda$CDM models, linear
perturbation theory provides good accuracy up to scales of  $k \sim 0.1 h \, {\rm Mpc}^{-1}$, whereas dark matter-only N-body
simulations provide good results up to $k \sim 0.5 h \, {\rm
  Mpc}^{-1}$, for larger values of $k$ it is necessary to
include  baryonic physics in the simulations \cite{Schneider:2015yka}.

For the SFDM model, we are able to describe the linear PS
at scales that can be considered in the semi-linear regime without any
by-hand approximation. A correct treatment of the linear and semi-linear regime impacts the study
of the structure formation process in two aspects: on setting up the
initial conditions for cosmological simulations, and on the prediction
(as a first approach) of observables about large and small scale
structure. 

\begin{figure*}[t!]

\includegraphics[width=0.48\textwidth]{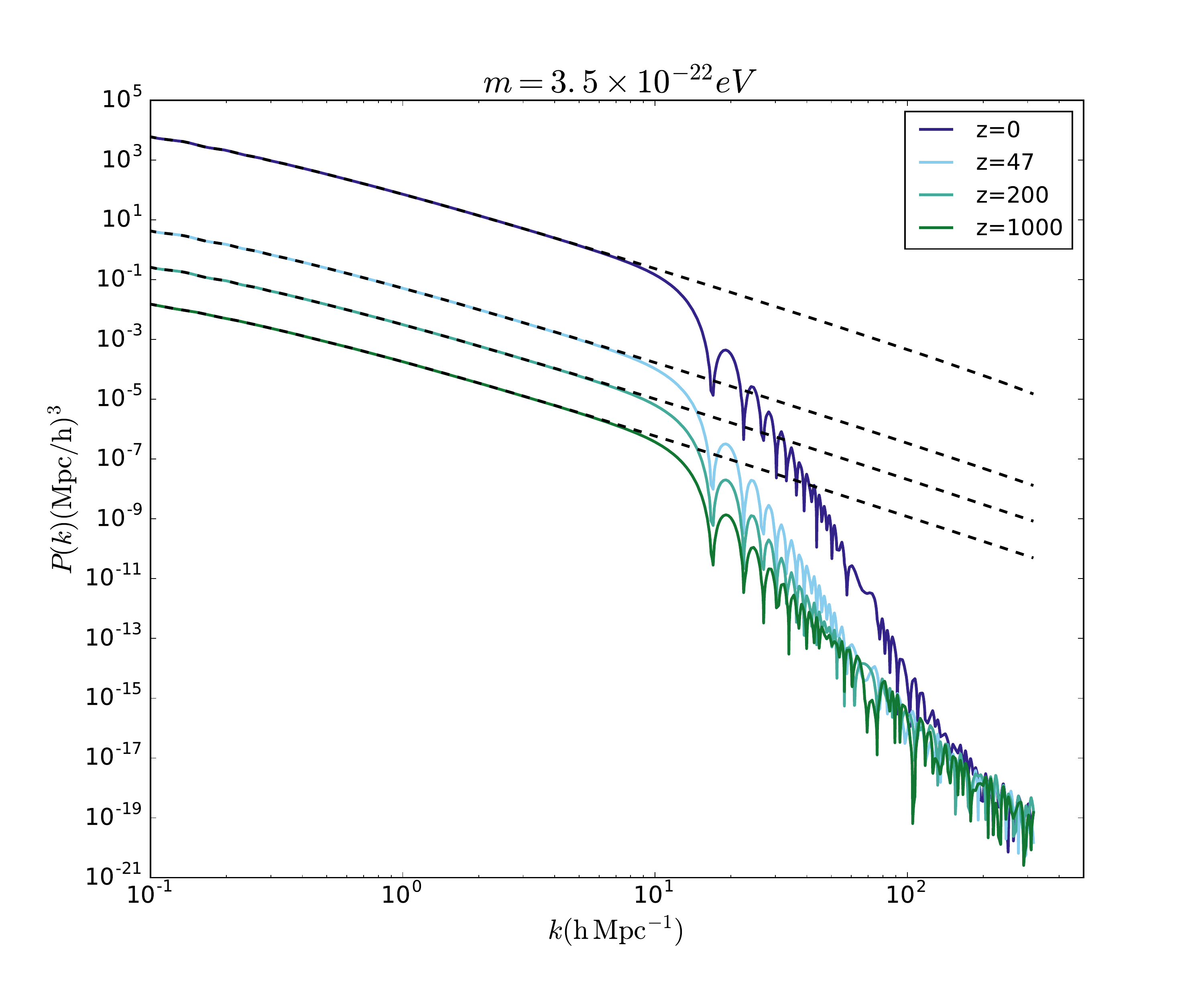}
\includegraphics[width=0.48\textwidth]{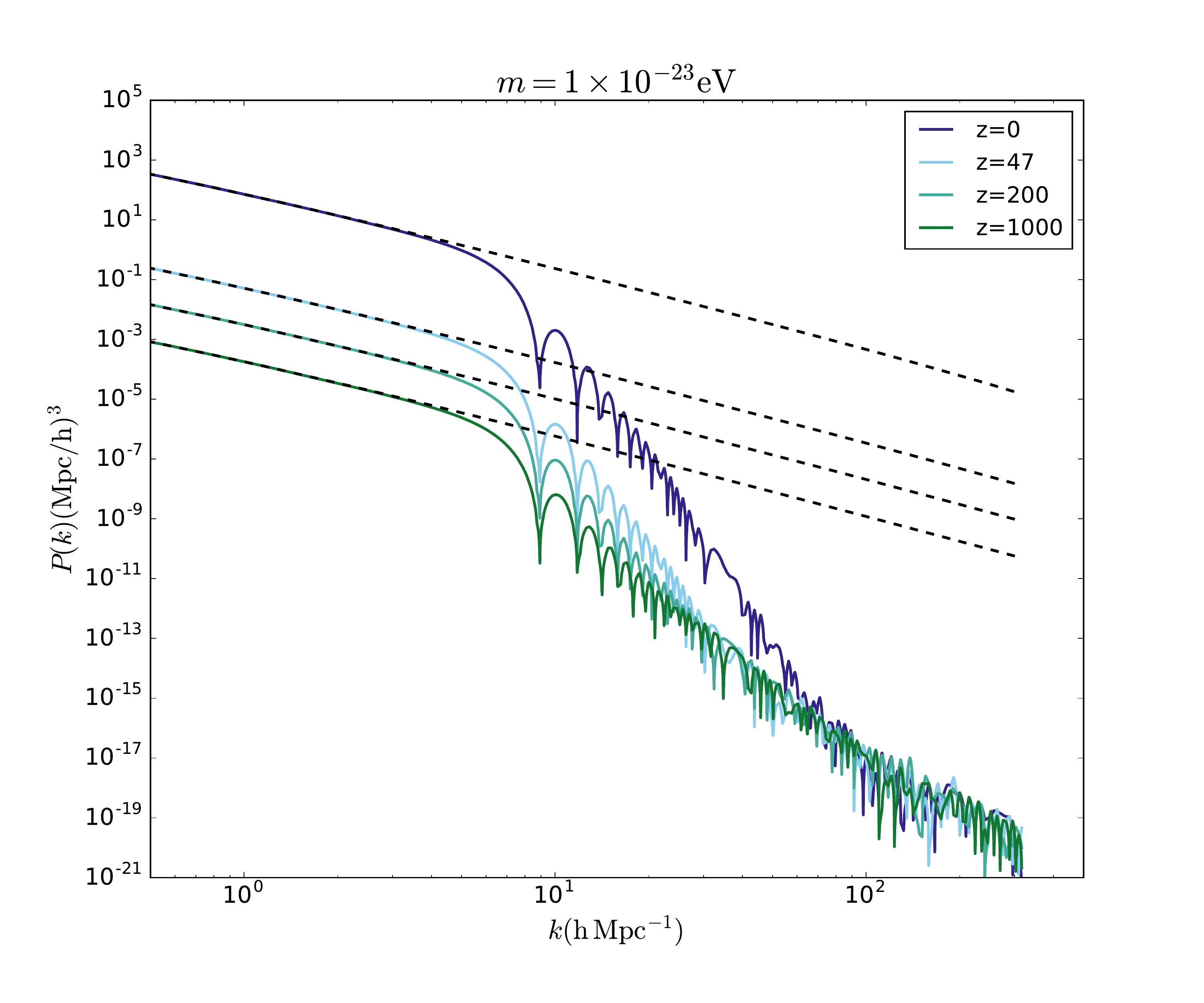}
\caption{\label{fig:PK_highz}  Mass PS for two values of
  the scalar field mass that resemble the initial condition of the
  cosmological simulation presented in Ref.~\cite{Schive:2014dra} (see
  figure S2 of that paper). Although we cannot make a precise
  comparison, the amplitude of the first oscillations seem to be in
  good agreement, but for lager values of $k$ the amplitude of the PS
  differ noticeably even by 9 orders of magnitude, which means that
  perturbations are much more suppressed than the estimation made
  in\cite{Schive:2014dra}. This difference might be important only if
  there is enough resolution in the simulations to include such
  $k$-values}
  
\end{figure*}

\begin{figure}[tp!]
  \includegraphics[width=\textwidth]{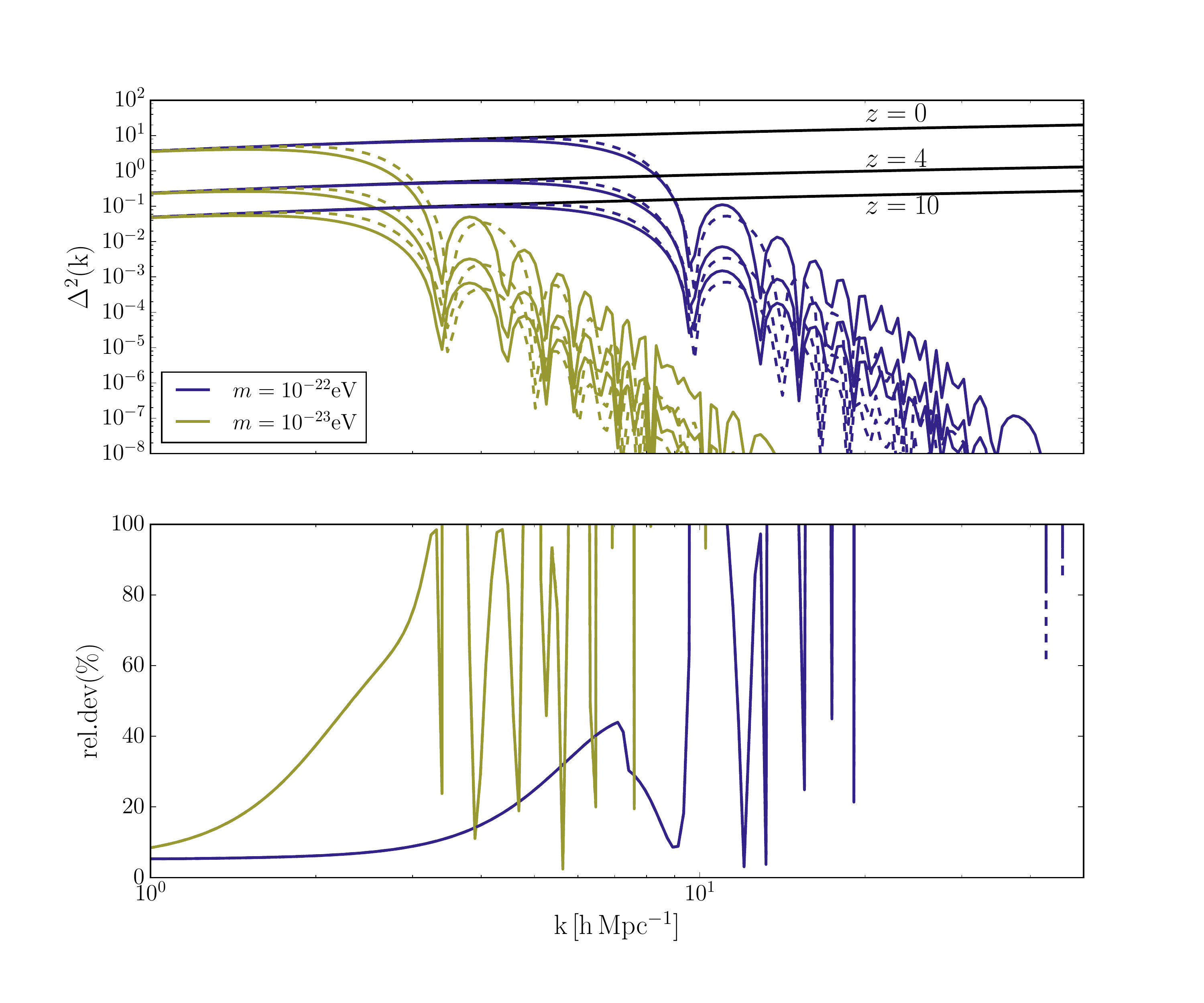}
  \caption{\label{fig:PK_Pcut_comp} Dimensionless PS
    at low and medium redshifts for two different values of the scalar
    field mass, and their comparison with the approximation in
    Eq.~\ref{eq:psShive}. In the bottom panel we show the error between
    the two approaches, which is noticeably larger for smaller values of
    the scalar field mass. We only show here the oscillatory range in
    the PS, see Fig.~\ref{fig:PK_diffz} for the differences
    at small values of $k$.}
\end{figure} 

Initial conditions for structure formation simulations are set up at
high redshift using the predicted PS for a given cosmology. In the
case of SFDM, two approaches have been so far considered for the
task. The first one starts a cosmological simulation at a very high
redshift $z\approx 1000$, when DM is just to dominate the expansion of
the Universe. For this case, we compare the PS used to set up the
initial conditions in\cite{Schive:2014dra}, to the one calculated from
our amended CLASS code. The resultant PS are shown in
Fig.~\ref{fig:PK_highz}, for two different values of the SF mass at
different redshifts: $z=0,47,200,1000$. A comparison with figure S2
in\cite{Schive:2014dra}, we note that for the two versions of the PS
the amplitude of the first oscillation matches quite well, but for
smaller scales the amplitudes seem to be overestimated in the
simulation, or likewise underestimated by us. At this point, though,
it is not possible to tell which of the approaches is the correct one,
since the CMB codes used are different: CMBFAST was used
in\cite{Schive:2014dra}, which is a code not longer supported by their
developers. However, the differences, mostly present at the largest
values of $k$, might be important only if the resolution in the
simulation is of the order of the Jeans scale.

The second approach is to use Eq.~\ref{eq:psShive} to just re-scale the
$\Lambda$CDM PS. In Fig.~\ref{fig:PK_Pcut_comp}, we show the PS
obtained from our modified CLASS code (solid lines) and the re-scaled
$\Lambda$CDM one (dashed lines), for two SFDM masses and for different
redshifts. Relative differences between the two PS are redshift
independent for the scales shown, but for the first oscillation they
can be as large as $40\%$ if $m=10^{-22}eV$ , and almost as $100\%$
if $m=10^{-23}eV$. For the values of $k$ in the oscillatory regime,
the differences can be at least of $20\%$. It seems then that using
the approximated PS of\ref{eq:psShive} could be introducing a
significant error, even though it predicts correctly the scale at
which the PS starts to oscillate. 

\begin{figure}[htp!]
\includegraphics[width=\textwidth]{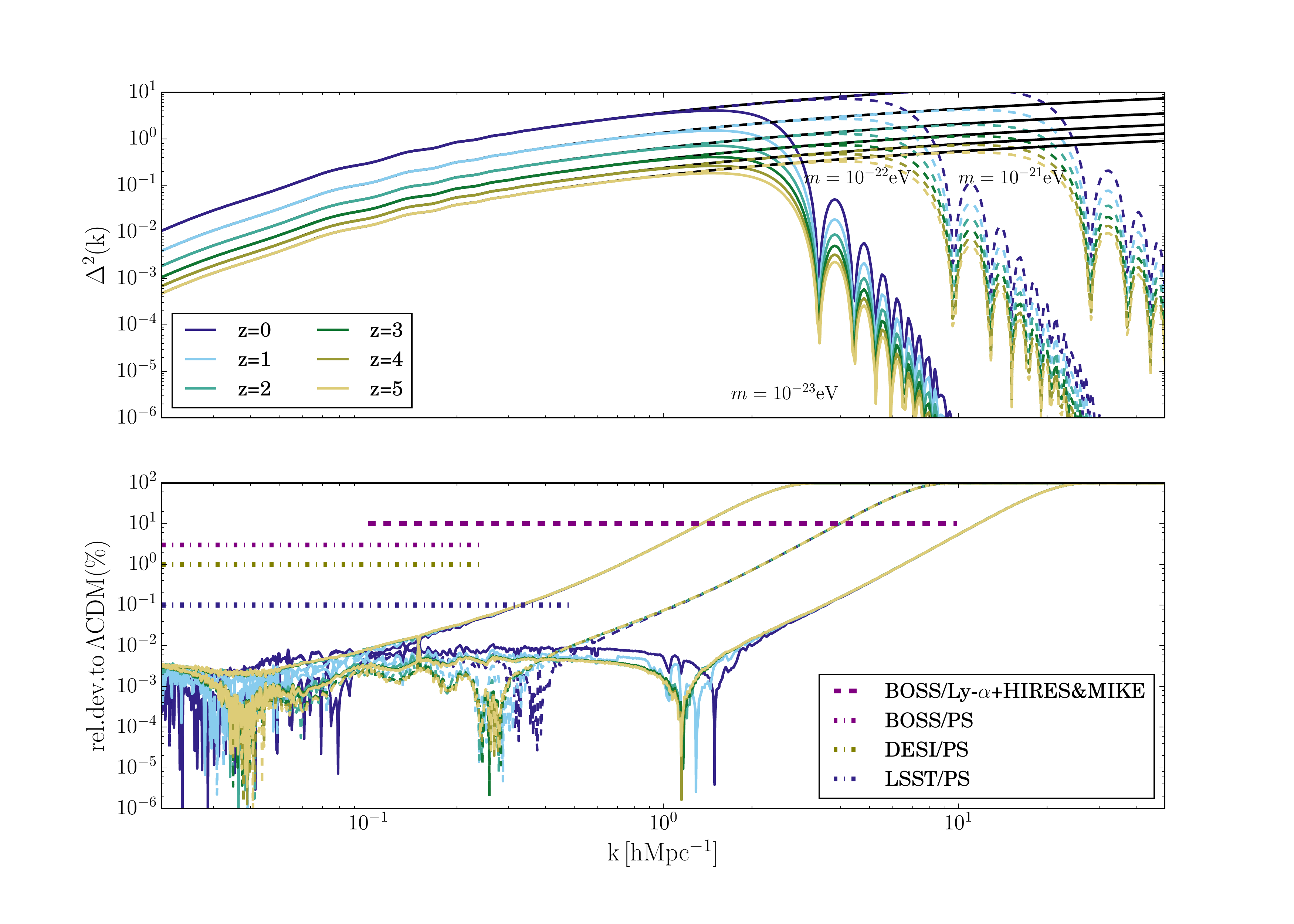}
\caption{\label{fig:PK_diffz} The same as in
  Fig.~\ref{fig:PK_Pcut_comp} but for the $k$-range that is usually
  tested with data from large scale galaxy and quasar surveys. A
  simple comparison shows that future LSST PS observations alone could
  place strong constraints in the mass range $m \gtrsim 10^{-23} \rm
  {eV} $.  DESI and LSST in combination with other observables could constrain the $10^{-23} -10^{-22} \rm
  {eV}$  range .}
  \end{figure}

On the other hand, for observables lying at high, medium, or low
redshifts, one could use the approximate form of the PS given by Eq.~\ref{eq:psShive}, as used in \cite{Schive:2014dra}, or
the numerical output from our amended version of CLASS. In 
Fig.~\ref{fig:PK_diffz} we  compare our resultant PS  to the LCDM one, and take the difference as an estimate of the precision needed in an observation in order to constrain the model.  SFDM models with masses larger
than $10^{-22}\, {\rm eV}$ seems not to be discriminated from the $\Lambda$CDM model, by the PS shape
obtained from current and next generation of galaxy surveys (BOSS/PS
and DESI/PS). However, the region in between $10^{-23} -10^{-22} {\rm
  eV}$ could be proved by the Lyma-$\alpha$ forest data
(BOSS/Ly-$\alpha$ and DESI/Ly-$\alpha$) only if information for larger
$k$ is included, such as the data from HIRES and MIKE, as was used in
\cite{Viel:2013apy}. Notice that for Warm Dark Matter (WDM)  models there is a strong
constraint on the suppression of the linear matter power spectrum at
(non-linear) scales corresponding to $k=10 \, h {\rm Mpc}^{-1}$, which
should not deviate more than $10\%$ from that of $\Lambda$CDM
model. For the SFDM this  implies a strong constraint up to masses of $\sim 10^{-21} {\rm eV}$. However one should take this constraint with care because it lays at highly no-linear scales,  and it could strongly depend on the non-linear evolution of the SFDM perturbations in the presence of baryons.

\begin{figure}
\begin{center}
\includegraphics[width=0.7\textwidth]{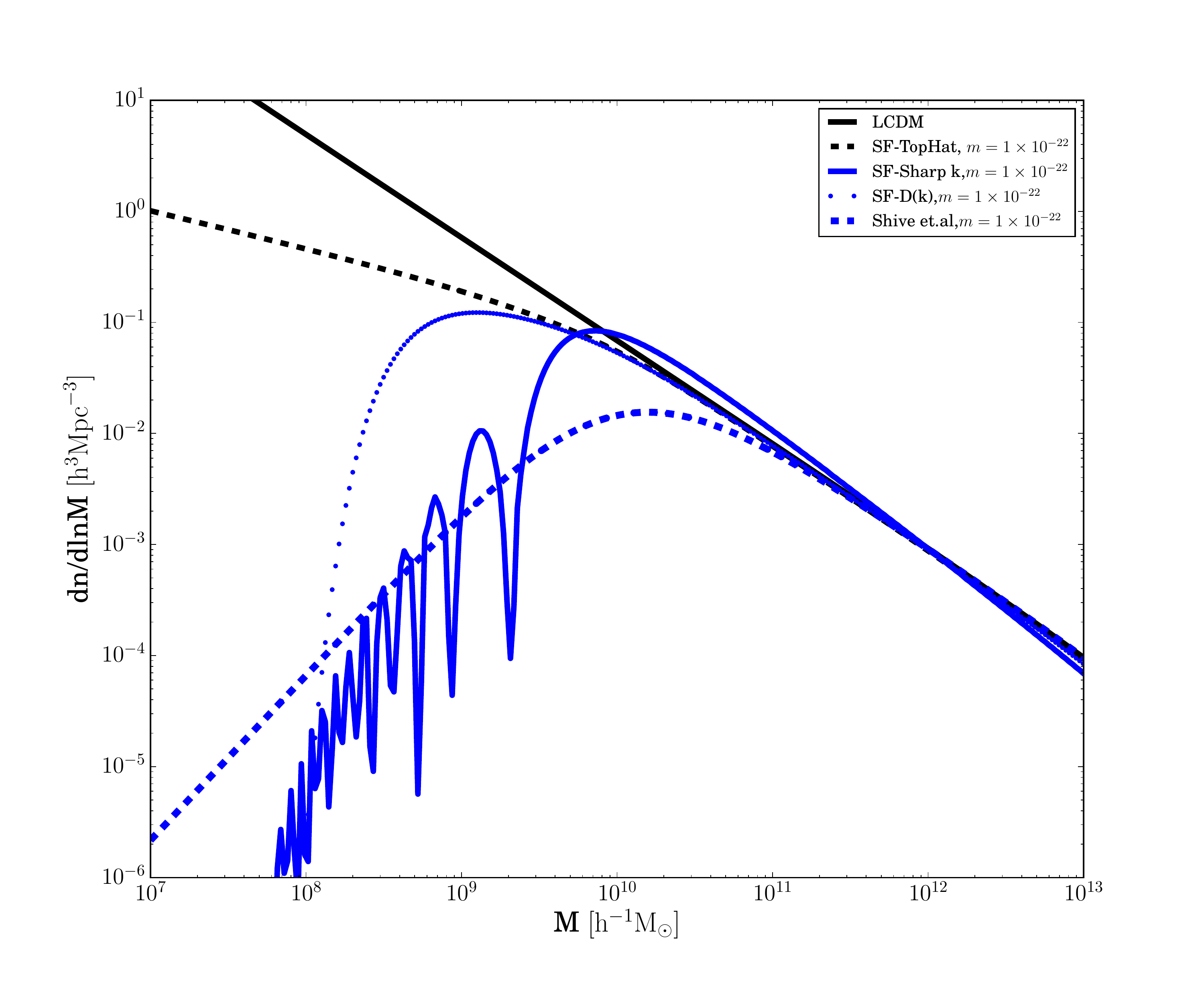}
\caption{\label{fig:mf1} Logarithmic derivative of the halo mass
  function for $\Lambda$CDM and $\Lambda SFDM$ (with $m = 10^{-22}
  {\rm eV}$). The cases depicted are the Top-hat (dashed curve),
  the Top-hat with a scale dependent growth function (dotted curve),
  the sharp-$k$ window function (solid curve), and the fit function
  inferred from the numerical simulations in\cite{Schive:2015kza}. The
  suppression of structures below $10^{10} {\rm h^{-1}M_\odot}$ is
  clear for all functions, although there are clear qualitative and
  quantitative differences among them. See the text for more details.}
\end{center}
\end{figure}

Notice,  also from  \ref{fig:PK_diffz},  that the LSST survey could,
at the very end of its $k $ coverage, set some constraints directly
from the PS at the semi-linear regime.  This is a very
simple estimation based on \cite{Zhan:2005rz} where they forecast the
power of constraining the primordial PS. If we relay on standard
assumptions about the primordial PS then constraints could be set SFDM
PS. A detailed forecast analysis will be done elsewhere, however it is
clear it will not be easy to discriminate between the effect of a
modification to the primordial  PS or modifications due to SFDM. The
information from other observables will be necessary. LSST will
probably set more constraints through weak and strong lensing, at
smaller scales (larger $k$'s), but such smaller scales are dominated
by the non-linear evolution of perturbations\cite{2009arXiv0912.0201L}.

By having  the mass PS at semi-linear scales we can make
rough predictions about different observables. One of the most
interesting quantities to compute is the halo mass function, which can
give us information about the fraction of the density that have
collapsed into self-gravitating structures of a given mass. It is also
the base function to have an estimation of the number of structures of
a certain mass that can be used to set constraints on cosmological
parameters through the evolution of the number of
clusters\cite{2013MNRAS.431.1866R}, or to infer properties about the
dark matter through the amount of expected Milky Way satellites
\cite{Schneider:2013ria}. Even though for these observations the
semi-analytic  mass function will only be an approximation since it includes non-linear physics only approximately . For observations at large
redshift it could give more reliable information \cite{Bozek:2014uqa}.

Previous studies of structure formation with small-scale suppression, particularly for WDM, have shown that the Top-hat
model fails in capturing the behavior of the cut-off in the power
spectrum, and how it translates into an almost complete suppression of
the small scale structure. In contrast, the sharp-$k$ window function
seems to better reproduce the mass function from simulations. Some
examples of these effects were shown
in\cite{2013MNRAS.428.1774B,Schneider:2014rda,Buckley:2014hja}. The
explanation provided in\cite{Schneider:2014rda} is that with a
sharp-$k$ filter the halo mass function depends on the shape of the
PS for any radius, while for a Top-hat filter it  becomes
completely insensitive to the PS shape  as soon as
the latter decreases faster than $k^3$. More recently, the simulations
for the SFDM model done by \cite{Schive:2015kza} also showed the
presence of a turn around in the halo mass function that seems to be
better described by a sharp-$k$ function than by the standard Top-hat
one. 
In Fig.~\ref{fig:mf1}, we show the halo mass function of SFDM computed
as in\cite{Schneider:2014rda,Schneider:2013ria}, where a semi-analytic
approach based on the original Press-Schechter methodology is
presented in the context of WDM. We used a Top-hat window function,
and a sharp-$k$ function, for a scalar field mass of $m=10^{-22} {\rm
  eV}$. We also computed the mass function using a mass dependent
barrier following \cite{Hlozek:2014lca} (see also a recent
implementation in\cite{Marsh:2016vgj}). We show for comparison the
halo mass function fit found from the numerical simulations
in\cite{Schive:2015kza}. We can note that the small mass scales are
very sensitive to the choice of the window function. Qualitatively,
the sharp cut-off in the PS	 translates into a small
suppression in the mass function if a Top-hat function is used, while
it translates into a more accute cut-off for the sharp-$k$ function
and the scale dependent growth factor. Since for the mass of our
choice the cut-off happens at mass scales smaller than $10^{10}$ Solar
masses, we still expect it to be consistent with current numbers of
observed satellite galaxies in the local group. The cut-off mass we
find with the sharp-$k$ window function captures well the result from
the numerical simulations in terms of the scale at which the turn is
happening, although there is a significant difference in the amplitude
for the parameters we used. Note that we set the the sharp-$k$
function to match $\Lambda$CDM at large masses. 

Finally, it is worth to mention that both window functions, Top-hat or
sharp-$k$, need the input from numerical simulations of structure
formation for a reliable computation of the halo mass function. The
sharp-$k$ function needs it to set the mass scale at the turn-around,
while the Top-hat needs it to set the $q$-value, which formally is a
free parameter, in the first-crossing
distribution. In\cite{Schneider:2014rda}, the mass at the PS turn
around was fixed using warm dark matter simulations, and it was argued
that this should be independent of the dark matter model, as soon as
the cut-off is present. We consider it should be fixed by simulations
of scalar field dark matter in order to not introduce a potential
source of error. From the comparison in Fig.~\ref{fig:mf1} among the different window functions and  the  functional form proposed  in\cite{Schive:2015kza}, it seems that the sharp-$k$ window function, is the one that is capturing better the non-linear effects, found in the simulations. However in such figure we just set the free
parameters in order to have a reasonable match of the halo mass
function to the $\Lambda$CDM one at large mass scales. Some work about
how the differences in the halo mass function propagate to the number
of satellites in a Milky Way type galaxy can be found
in\cite{Marsh:2013ywa}.

\section{Conclusions \label{sec:conclusions}}
We have presented a new method to transform and solve numerically the
equations of motion of a scalar field as a dark matter component. This
method takes advantage of the explicit symmetries of the coupled
Einstein-Klein-Gordon system of equations, and allow a clear
understanding of the underlying physical solutions of the scalar
field. The method is also suitable for numerical purposes, and deals
correctly with the rapid oscillations of the field without any
pre-handling of the equations of motion.

We showed that the background dynamics of SFDM is very simple in terms
of the new dynamical variables, and even eases its numerical
implementation in CMB codes, as exemplified with the results obtained
from CLASS. Moreover, the transformation of variables used for the
background case was easily adapted for the evolution of linear
perturbations in the scalar field, and there was no need to rely on
any similarities with standard fluid dynamics. Under this approach, we
were able to find semi-analytical solutions which show in a clear way
the physical effects acting in the growing or suppression of linear
perturbations.  The amended version of CLASS will be used in
  future studies to determine the direct effects of scalar fields in
  different observables, specially on those that make the CMB spectrum
distinguishable from that of $\Lambda$CDM (see Fig.~\ref{fig:3}), for
both types of dark matter and dark energy fields. Another topic of
interest is the setting up of initial conditions of $N$-body and
wave-like simulations, so that one can incorporate constraints for
scalar field models from linear and non-linear perturbations.

Our results indicate that the growing of perturbations in SFDM models
for a given wave number $k$ requires of two conditions: a) an early
onset of field oscillations, preferably much earlier than the time of
radiation-matter equality, and b) a negligible ratio $k/k_J$. Not
surprisingly, both conditions depend upon the only free physical
parameter which is the mass of the SFDM particle, and actually the two
conditions are simultaneously enhanced (suppressed) in the case of
large (small) values of $m$ just because both angular variables
$\theta$ and $\vartheta$ evolve similarly throughout.

As for condition b), we must recall first that the scales that have
been affected at some degree by the Jeans wavenumber $k_J$, throughout
the evolution of the Universe, are those for which initially
$k/k_{J,eq} > 1$ with $k_{J,eq} = \Omega^{1/4}_{r0} k_{J0}$, see
Eq.~\eqref{eq:23b}. From this, we can infer that suppression of power,
as compared to the standard CDM case, happens in scales:
\begin{equation}
  \frac{k}{\rm Mpc} > 0.68 \, \Omega^{-1/4}_{r0} \left( \frac{m}{10^{-23}
      {\rm eV}} \right)^{1/2} \left( \frac{H}{\rm km \, s^{-1} \,
      Mpc^{-1}} \right)^{1/2} \, . \label{eq:27b}
\end{equation}
This suggests, for instance, that for $m \geq 10^{-24} {\rm eV}$ even
the slightest suppression of perturbations will lie at scales $k > 1
\, {\rm Mpc}^{-1}$, an estimation that is in agreement with the
numerical results shown in Fig.~\ref{fig:3}. In summary, for $m \geq
10^{-24} {\rm eV}$ the main source of suppression of SFDM linear
perturbations is the Jeans length $k_J$, whereas for smaller values of
$m$ linear perturbations are also suppressed by the late appearance of
the scalar field oscillations in the background.

We have recovered the results for SFDM perturbations that have been
already reported in the literature under the fluid-approximation of
the equations of motion for
perturbations\cite{Park:2012ru,Hlozek:2014lca,Marsh:2010wq}. However,
we would like to emphasize that our method does not have a direct
translation to the fluid-approximation, because some fluid quantities
cannot be well defined in terms of our variables. An example here is
the sound speed of linear perturbations, which in our variables would
be written as:
\begin{equation}
  \label{eq:46}
  c^2_s = \frac{\delta p_\phi}{\delta \rho_\phi} =
  \frac{\sin(\theta +\tilde{\vartheta}/2)}{\sin(\tilde{\vartheta}/2)}
  = \sin \theta \tan (\tilde{\vartheta}/2) + \cos \theta \simeq
  \sin(2mt) \tan (\tilde{\vartheta}/2) + \cos (2mt) \, ,
\end{equation}
where the last expression on the rhs is the approximate one after the
onset of the field oscillations, see also Eqs.~\eqref{eq:26}. From our
perspective, $c^2_s$ becomes a rapidly-oscillating function, and then
our cut-off procedure would dictate that $c^2_s \to 0$ for both large
and small scales. This is at variance with the fluid approximation
used in previous papers\cite{Park:2012ru,Hlozek:2014lca,Marsh:2010wq},
and seems also to be an early indication that a direct translation to
the fluid notation may not be straightforward. For this reason, the
connection of our variables to those of the fluid approximation is a
topic that deserves further investigation, but we leave this issue for
a future work.

Finally, we have explored some differences that appear in the mass
PS with respect to previous results published in the
literature. The wavenumber at with which the cut-off of the power
spectrum appears is consistent with previously reported results \cite{Marsh:2013ywa,Schive:2015kza}, but the
suppression of power is not as sharp; this small differences can be
significant for the determination of the mass function of small scale
structure. The mass of the scalar field needs to be larger that
$10^{-23} {\rm eV}$ in order to be compatible with current power
spectrum measurements. Future LSST survey will probably be able to
prove the region $m\gtrsim 10^{-23} {\rm eV}$ (competitive to e.g.\cite{ Bozek:2014uqa} from reionization or to \cite{Khmelnitsky:2013lxt} from pulsar timing).
A further analysis of these constraints will be presented elsewhere. 

\begin{acknowledgments}
We want to thank Francisco Linares for a careful reading of the
manuscript. We are grateful to an anonymous referee for enlightening
comments and suggestions. LAU-L wishes to thank Andrew Liddle and the
Royal Observatory, Edinburgh, for their kind hospitality in a fruitful
sabbatical stay, where part of this work was done. AXGM
acknowledges support from C\'atedras CONACYT and UCMEXUS-CONACYT
collaborative proyect funding. This work was partially supported by
PRODEP, DAIP-UGTO research grant 732/2016 and  878/2016, PIFI, CONACyT M\'exico under
grants 232893 (sabbatical), 167335, 179881 and Fronteras 281,
Fundaci\'on Marcos Moshinsky, and the Instituto Avanzado de
Cosmolog\'ia (IAC) collaboration.
\end{acknowledgments}

\appendix
\section{Cut-off procedure for rapidly oscillating
  terms \label{sec:cut-proc-rapidly}}
  Even though the numerical solutions seem to be well behaved after
  the cut-off implemented on the rapidly oscillating functions that
  appear in the equations of motion, the question remains of whether
  the cut-off is erasing any important part of the expected
  solution. As we will show now, the full solutions of different SF
  quantities can be decomposed in a monotone solution plus rapidly
  oscillating terms that quickly disappear at late times after the
  onset of rapid oscillations.

\subsection{Background variables}
\label{sec:background-variables}
 The simplest case is that of variable $\theta$ itself, for which
  there is an exact solution. From Sec.~\ref{sec:early-late-time}, we
  learned that Eq.~\eqref{eq:4a} can be written in terms of the cosmic
  time as:
\begin{equation}
\dot{\theta} = 2m -3 H \sin \theta \, . \label{eq:47}
\end{equation}
It was shown that the second term on the rhs of Eq.~\eqref{eq:47}
becomes negligible after the onset of rapid oscillations, and then the
dominant solution at late times is $\theta(t)= 2mt$. The latter is
also the solution that is obtained once we apply the cut-off on the
rapidly oscillating term on the rhs of Eq.~\eqref{eq:47} at
$\theta(t_\star) =\theta_\star$.

This suggests that an approximate solution after the cut-off can be
obtained by the substitution of $2mt$ on the rhs of Eq.~\eqref{eq:47},
and then:
\begin{subequations}
\begin{equation}
  \label{eq:48}
  \theta(t) = 2mt -3 \int^t_{t_\star} H \sin(2mt) \, dt \, .
\end{equation}
The Hubble parameter is given by $H=p/t$, where $p=1/2$ ($p=2/3$) for
radiation (matter) domination, so that Eq.~\eqref{eq:48} can be
rewritten as:
\begin{equation}
  \label{eq:51}
   \theta(t) = 2mt -3p
   \int^{\theta}_{\theta_\star} \frac{\sin x}{x} \, dx =
   2mt - 3p \left[ {\rm Si}(\theta) - {\rm
     Si}(\theta_\star) \right] \, ,
\end{equation}
\end{subequations}
where ${\rm Si}(x)$ is the sine integral. Thus, the solution of
$\theta(t)$ consists of the linearly-growing term $2mt$, plus a
rapidly oscillating function with a decaying amplitude. From the
properties of the sine integral, we find that $[ {\rm Si}(\theta) -
{\rm Si}(\theta_\star)]/\theta < 8.6 \times 10^{-5}$ if $\theta >
\hat{\theta}_\star = 10^2$. Thus, we can safely say that the cut-off
procedure applied to Eq.~\eqref{eq:4a} is just eliminating a
negligible part of the full solution of $\theta(t)$ at late times.

We now use the above result to obtain an approximate solution of
the density parameter after the the rapidly oscillating terms of are
cut-off. Eq.~\eqref{eq:4c} can be written as: 
\begin{equation} 
  \label{eq:29}
  d \left( \ln \Omega_\phi \right) = 3 w_{tot} \, dN + 3 H \cos
  \theta \, dt \simeq  3 w_{tot} \, dN + 3 p \frac{\cos \theta}{\theta}
  d\theta \, ,
\end{equation}
where we have assumed that $\theta = 2mt$ for $t >
t_\star$. Eq.~\eqref{eq:29} indicates the existence of two different
time scales in the evolution of $\Omega_\phi$: the small one
corresponding to rapid oscillations and represented here by $\theta$,
and the large one represented by the number of $e$-folds $N$. Upon
integration, we find that:
\begin{eqnarray}
  \Omega_\phi(t) &=& \Omega_{\phi \star} \exp \left[ 3 \int^N_{N_\star}
                     w_{tot} \, dx \right] \exp \left[ 3 p
                     \int^\theta_{\theta_\star} \frac{\cos x}{x} dx
                     \right] \, , \nonumber \\
                 &=& \Omega_{\phi \star} \exp \left[ 3 \int^N_{N_\star}
                     w_{tot} \, dx \right] \exp \left[ 3 p ({\rm
                     Ci}(\theta_\star) - {\rm Ci}(\theta))
                     \right] \, , \label{eq:30}
\end{eqnarray}
with ${\rm Ci(x)}$ being the cosine integral, and $\Omega_\ast = \Omega(t_\ast)$. As expected, $\Omega_\phi$ has a
steady evolution provided by the integral in the first exponential
term on the rsh of Eq.~\eqref{eq:30}, with some superimposed
oscillations that decay away very quickly, which are obtained from the
cosine integral in the last exponential term. Such oscillations in
$\Omega_\phi(t)$ can even be seen by eye, for example, in some of the
numerical solutions shown in Fig.~\ref{fig:1}.

The amended version of CLASS gives the same result for the density
parameter as that obtained directly from Eq.~\eqref{eq:4c} before the
cut-off is applied to the rapidly oscillating functions. After this
time, that is for $\theta > \theta_\star$, the solution provided by
CLASS simply is:
\begin{equation}
  \label{eq:77}
  \Omega^{\rm CLASS}_\phi(t) = \Omega_{\phi \star} \exp \left[ 3 \int^N_{N_\star}
    w_{tot} \, dx \right] \, .
\end{equation}
We can clearly see that the cut-off procedure is equivalent to putting
away the superimposed oscillations in the solution of
$\Omega_\phi$. A quick comparison between Eqs.~\eqref{eq:30}
and~\eqref{eq:77} shows that for $\theta > \theta_\ast$
\begin{equation}
  \label{eq:70}
  \frac{\Omega_\phi(t) - \Omega^{\rm CLASS}_\phi(t)}{\Omega_\phi(t)} < 3p \,
  {\rm Ci}(\theta_\star) \, ,
\end{equation}
and then the relative error between the true and the amended numerical
solutions depends solely on $\theta_\star$. For our empirical choice
of $\theta_\star = 10^2$, we find that ${\rm Ci}(10^2) \simeq -5.1
\times 10^{-3}$, which means that the relative error in the density
parameter calculated from CLASS with respect to the true solution
grows with time but is always less than $1$\%.

\subsection{Numerical examples}
\label{sec:numer-exampl-from}
\begin{figure*}[htp!]
\centering
\includegraphics[width=0.45\textwidth]{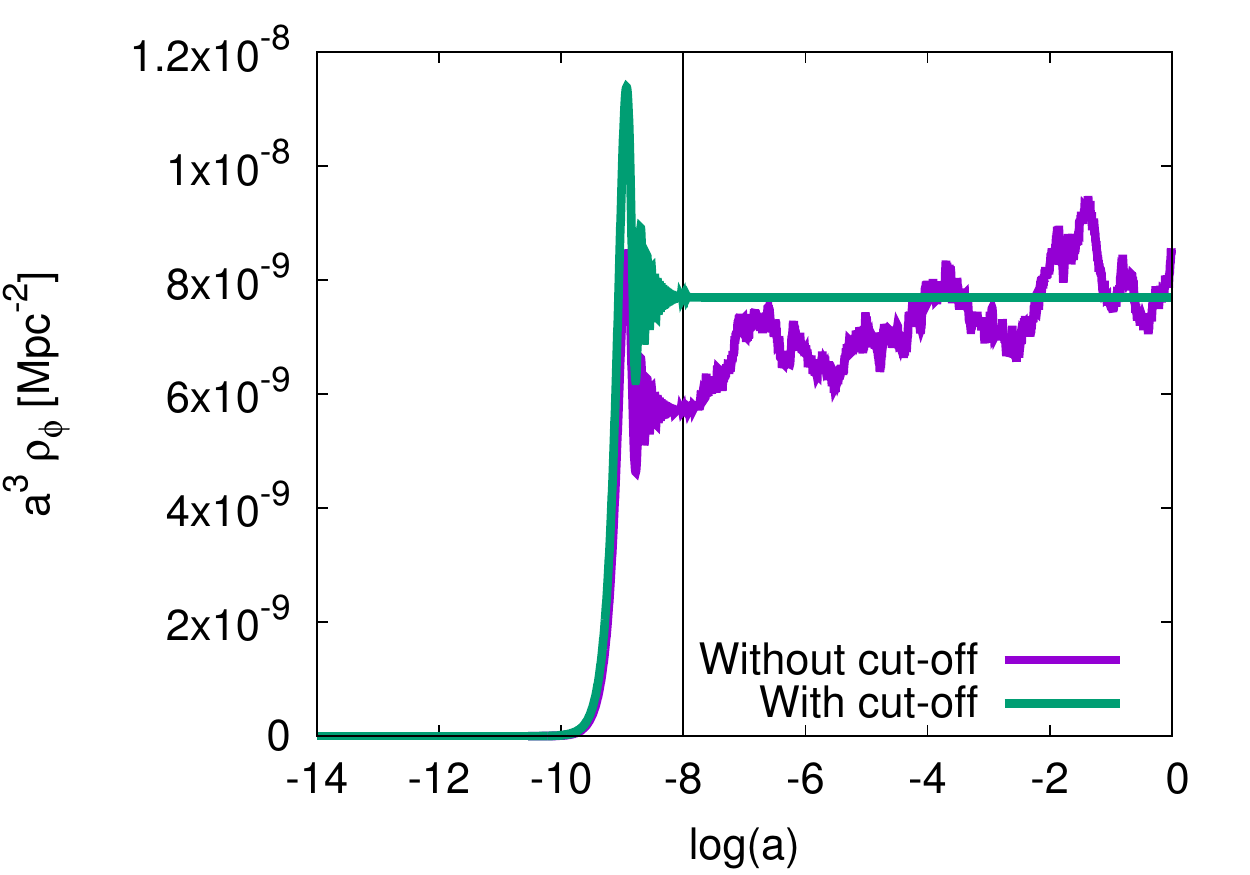}
\includegraphics[width=0.45\textwidth]{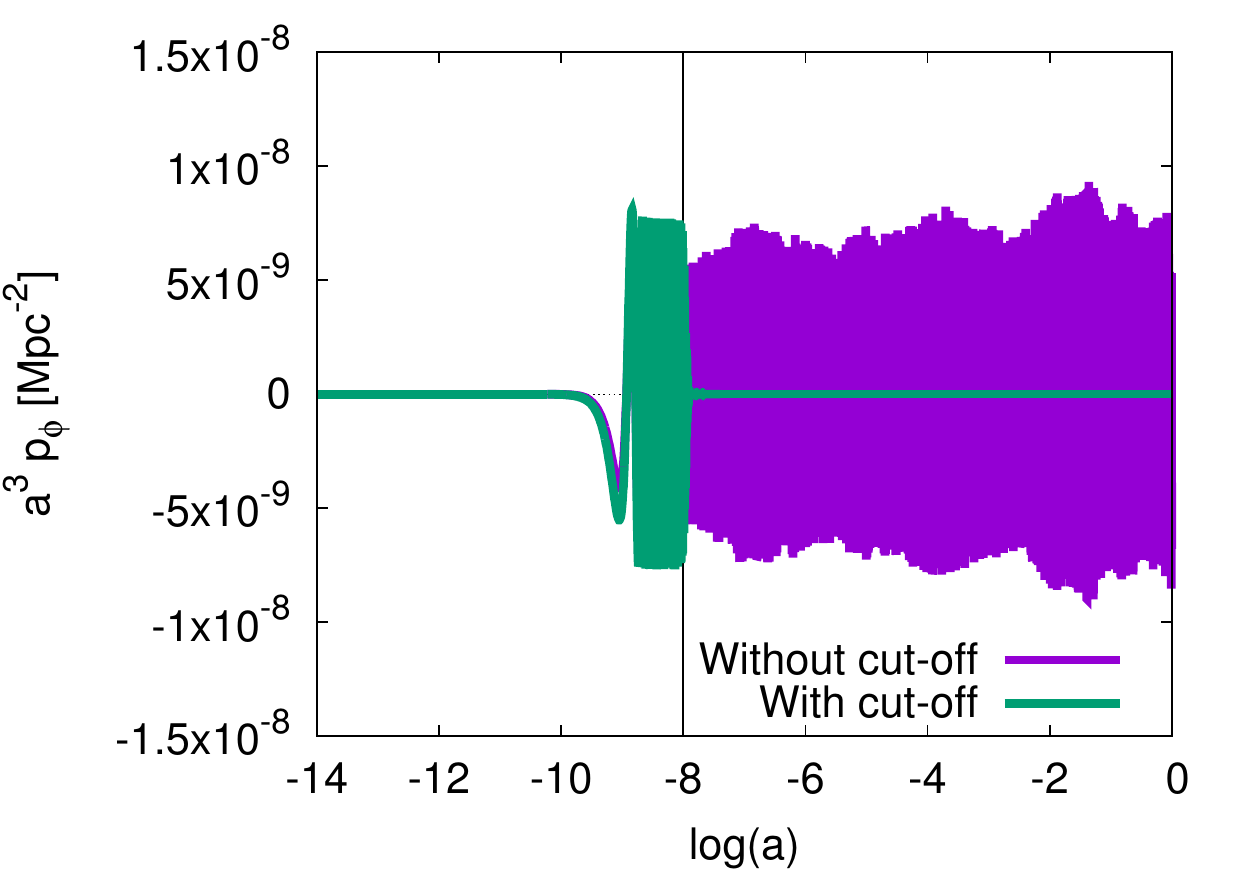}
\caption{\label{fig:2} (Left) The normalized energy density $a^3 \kappa^2
  \rho_\phi /3$ in CLASS units, for a comparison between the cut and
  uncut cases of high frequency oscillations. We notice that the CLASS
  equation solver cannot give accurate numerical results for $a >
  10^{-8}$, but the cut-off case gives a smooth transition to the
  correct solution at late times. In this case, the empirical choice
  $\theta_\star = 10^2$ means that the cut-off, indicated by the
  vertical line, is applied at $a = 10^{-8}$. (Right) The normalized
  pressure $a^3 \kappa^2 p_\phi/3$ for the cut and uncut cases. The
  cut-off in the oscillations is more drastic in this case, although
  it can be seen that the pressure oscillates a bit erratically around
  zero in the full numerical solution, which signals the limitations
  of CLASS when dealing with stiff systems of equations. See the text
  and Sec.~\ref{sec:modifications-class} for more details.}
\end{figure*}

In order to show the behavior of the numerical solutions under the
cut-off procedure implemented for the trigonometric functions through
Eq.~\eqref{eq:18}, we present in Fig.~\ref{fig:2} the cases of the
scalar field energy density $\rho_\phi$ and pressure $p_\phi$; the
examples correspond to the same Universes shown in Fig.~\ref{fig:1}
with a mass of $m = 10^{-22} {\rm eV}$. First of all, it can be seen
that the uncut case cannot be handled appropriately by the equation
solver within CLASS, and the evolution of the scalar field variables
appears as erratic after the onset of the rapid oscillations. One
undesirable consequence of this is that the present contribution of
the scalar field, as given by $\Omega_{\phi 0}$, cannot be determined
accurately enough from the given initial conditions discussed in Sec.~\ref{sec:initial-conditions-}.

The evolution with the cut-off implemented, on the other hand, is far
more stable, and, in the case of the energy density $\rho_\phi$, it
matches smoothly the expected behavior at late times described in
Secs.~\ref{sec:early-late-time} and~\ref{sec:background-variables}. In
contrast to the uncut case, the present scalar field contribution
$\Omega_{\phi 0}$ is given exactly as desired, and also determined
uniquely in terms of the initial conditions (see Eq.~\eqref{eq:19}). As
for the scalar field pressure $p_\phi$, see Eq.~\eqref{eq:17}, it is
drastically affected by the sudden cut-off of the oscillations in the
scalar field EoS $w_\phi$, but this does not have any major
consequences for the overall evolution of all other cosmological
variables. This again confirms that the cut-off procedure is a least
invasive modification of the numerical computations, and  then a
complementary one to the method of averaged quantities in the fluid approximation.

\subsection{Linear perturbations}
\label{sec:linear-perturbations}
 For the case of the perturbation variables, we can follow the
  same procedure as before but with some subtleties. To begin with, we
  first recall that after the onset of rapid oscillations $\theta =
  2mt$, see Secs.~\ref{sec:early-time-behavior}
  and~\ref{sec:background-variables}. Then, we write
  Eqs.~\eqref{eq:41} in terms of the cosmic time in the form:
\begin{subequations}
  \begin{equation}
    \label{eq:59}
    \dot{\mathbf{P}} = H \mathbf{P}^\prime = \left[ H \mathcal{M}_0(t)
      + \mathcal{M}_1(t) \right] \mathbf{P} + \frac{1}{2} h^\prime
    \left[ H \mathbf{N}_0 + \mathbf{N}_1(t) \right] \, ,
  \end{equation}
where $H$ is the Hubble parameter, a prime denotes derivative with
respect to the number of $e$-folds $N$, and
\begin{eqnarray}
  \label{eq:71}
  \mathbf{P} &=& \left[ 
    \begin{array}{c}
      e^\alpha \cos(\tilde{\vartheta}/2) \\
      e^\alpha \sin(\tilde{\vartheta}/2)
    \end{array} 
\right] \, , \quad \mathcal{M}_0(t) = \frac{\omega}{2} \left[
\begin{array}{cc}
0 & 1 \\
-1 & 0
\end{array}
\right] \, , \\
  \mathcal{M}_1(t) &=& - \frac{H}{2} \left[ 
\begin{array}{cc}
- \omega \sin(2mt) & 6 \sin(2mt) - \omega \cos(2mt) \\
- \omega \cos(2mt) & 6 \cos(2mt) + \omega \sin(2mt)
\end{array}
\right] \, , \\
\mathbf{N}_0 &=& \left[
\begin{array}{c}
0 \\
1
\end{array}
\right] \, , \quad \mathbf{N}_1(t) = H \left[ 
\begin{array}{c}
\sin(2mt) \\
-\cos(2mt)
\end{array}
\right] \, .
\end{eqnarray}
\end{subequations}
In order to understand the general behavior of the solutions, we will
again assume that $H=p/t = 2mp/(2mt) = 2mp/\theta$, and then we write Eq.~\eqref{eq:71} in the following differential form:
\begin{subequations}
  \label{eq:73}
\begin{equation}
  d \left[ e^{\mathcal{U}_0(N)} e^{\mathcal{U}_1(\theta)} \mathbf{P}
  \right] =  \frac{h^\prime}{2} e^{\mathcal{U}_0(N)}
  e^{\mathcal{U}_1(\theta)} \mathbf{N}_0 \, dN + \frac{h^\prime}{2}
  e^{\mathcal{U}_0(N)} e^{\mathcal{U}_1(\theta)} \mathbf{N}_1(\theta)
  \, d\theta \, , \label{eq:73a}
\end{equation}
where $dN= H \, dt$, and
\begin{equation}
  \mathcal{U}_0(t) = \int \mathcal{M}_0(N) \, dN \, , \quad
  \mathcal{U}_1(\theta) = \int \mathcal{M}_1(\theta) \, d\theta \, .   \label{eq:73b}
\end{equation}
\end{subequations}
As in the case of Eq.~\eqref{eq:29}, Eqs.~\eqref{eq:73} also indicate
the presence of the two different time scales in the evolution of
linear perturbations. In contrast to Eq.~\eqref{eq:73a}, the
differential equation that is solved by the amended version of CLASS
after the application of the cut-off on the rapidly oscillating terms
is:
\begin{subequations}
\begin{equation}
  \label{eq:78}
  d \left[ e^{\mathcal{U}_0(N)} \mathbf{P}^{\rm CLASS} \right] =
  \frac{h^\prime}{2} e^{\mathcal{U}_0(N)} \mathbf{N}_0 \, dN \, ,
\end{equation}
which then deals only with the time scale represented by $N$. Its
solution is: 
\begin{equation}
  \label{eq:27}
  \mathbf{P}^{\rm CLASS}(N) = e^{-\mathcal{U}_0(N)} \mathbf{P}_\star +
  \frac{1}{2}\int^N_{N_\star} h^\prime e^{[\mathcal{U}_0(\hat{N}) -
    \mathcal{U}_0(N)]} \mathbf{N}_0 \, d\hat{N} \, ,
\end{equation}
\end{subequations}
where $\mathbf{P}_\star = \mathbf{P}(t_\ast)$. Actually, it can be
easily verified that Eq.~\eqref{eq:78} is equivalent to
Eq.~\eqref{eq:31}, whereas we recover Eq.~\eqref{eq:32} from
Eq.~\eqref{eq:27} under the assumption that $\omega = {\rm const}$.

Like in the case of the background variables, we only need to show that
the integration of the rapidly oscillating terms in the general
solution of~\eqref{eq:73a} have become irrelevant by the time the
cut-off procedure is applied to the equations of motion. We first take
a look at Eqs.~\eqref{eq:73b}. The integration of the matrix function
$\mathcal{M}_1$ involves the use of the following cosine and sine
integrals:
\begin{subequations}
  \label{eq:74}
\begin{eqnarray}
  \int^t_{t_\star} H \cos(2mt) \, dt &=& p \int^\theta_{\theta_\star}
  \frac{\cos x}{x} dx = p [{\rm Ci}(\theta_\star) - {\rm Ci}(\theta)]
                                         \, ,   \label{eq:74a} \\
  \int^t_{t_\star} H \sin(2mt) \, dt &=& p \int^\theta_{\theta_\star}
  \frac{\sin x}{x} dx = p [{\rm Si}(\theta) - {\rm Si}(\theta_\star)]
                                         \, ,   \label{eq:74b} \\
  \int^t_{t_\star} \omega H \cos(2mt) \, dt &=& \frac{k^2}{2m}
                                                \int^\theta_{\theta_\star}
                                                \frac{\cos x}{a^2(x)} dx
                                                = \frac{\omega_\star}{2} [{\rm
                                                Ci}(\theta_\star) -
                                                {\rm Ci}(\theta)] \,
                                                , \label{eq:74c}
\end{eqnarray}
\end{subequations}
where the last integral~\eqref{eq:74c} was made under the assumption
of radiation domination only, and then $\omega_\star$ is the frequency
$\omega$ at the time $t_\star$. As argued before for
Eqs.~\eqref{eq:51} and~\eqref{eq:77}, because of our choice
$\theta_\star = 10^2$ and the presence of the cosine and sine
integrals in Eqs.~\eqref{eq:74}, we can see that the elements of the
matrix $\mathcal{U}_1$ are already small enough (of the order of
$10^{-2})$ when the cut-off procedure is applied. 

This suggests that we can safely replace $e^{\mathcal{U}_1(\theta)}
\simeq \mathcal{I} + \mathcal{O}(10^{-2})$, where $\mathcal{I}$ is the
unity matrix, and then Eq.~\eqref{eq:73a} can be written as:
\begin{equation}
  \label{eq:72}
  \mathbf{P}(N,\theta) \simeq \mathbf{P}^{\rm CLASS}(N) + \frac{1}{2}
  e^{-\mathcal{U}_0(N)} \int^\theta_{\theta_\star} h^\prime
  e^{\mathcal{U}_0(N)} \mathbf{N}_1(\theta) \, d\theta \, , \quad
  \mathbf{N}_1(\theta) = p \left[ 
\begin{array}{c}
\sin(\theta)/\theta \\
-\cos(\theta)/\theta
\end{array}
\right] \, ,
\end{equation}
where $\mathbf{P}^{\rm CLASS}(N)$ is the solution in Eq.~\eqref{eq:27}.
Even though the second term on the rhs of Eq.~\eqref{eq:72} still is a
rapidly oscillating function, it cannot be discarded as simply as
others due to the presence of $h^\prime$.

The last integral in Eq.~\eqref{eq:72} can be integrated iteratively
by parts, and in such case it can be proved that if we retain the
dominant term in the integrations we obtain\footnote{Interestingly
  enough, the solution~\eqref{eq:72} can be obtained by means of a
  direct integration of Eq.~\eqref{eq:73a} if $N$ and $\theta$ are
  considered \emph{independent} variables. This would be a reasonable
  short-cut to get  Eq.~\eqref{eq:73a} given that the variables
  indeed represent two completely different time scales in the general
  solution.}:
\begin{equation}
  \label{eq:79}
  \mathbf{P}(N,\theta) \simeq \mathbf{P}^{\rm CLASS}(N) + \frac{
    h^\prime}{2} \mathbf{N}_2(\theta) \, , \quad \mathbf{N}_2(\theta)
  = p \left[ 
\begin{array}{c}
{\rm Si}(\theta) - {\rm Si}(\theta_\star) \\
{\rm Ci}(\theta) - {\rm Ci}(\theta_\star)
\end{array}
\right] \, .
\end{equation}
We can apply here the same kind of trick we used in
Sec.~\ref{sec:gener-solut-pert}, where we inferred that for the
growing case: $\mathbf{P}^{\rm CLASS} \sim h^\prime/2$, see
Eq.~\eqref{eq:32}. Thus, the second term on the rhs of
Eq.~\eqref{eq:79} adds a negligible oscillating function to the
general solution; in other words, we have just found that
$\mathbf{P}(N,\theta) \simeq \mathbf{P}^{CLASS}(N) (1 +
\mathcal{O}(10^{-2}))$, which is the same level of accuracy attained
for the background variables.

\bibliography{oscillating2}

\end{document}